\documentclass[aps,prb,preprint,groupedaddress]{revtex4-1}
\usepackage{amsmath,graphicx,siunitx}
\usepackage[LGRx,T1]{fontenc} 
\usepackage[utf8]{inputenc}
\usepackage[greek,english]{babel}

\begin{document}


\title{Photochemical Upconversion Light Emitting Diode (LED):  Theory of Triplet Annihilation Enhanced by a Cavity}


\author{Laszlo Frazer}
\affiliation{ARC Centre of Excellence in Exciton Science, School of Chemistry, Monash University, Victoria, Australia}


\date{\today}

\begin{abstract}
Artificial lighting is a widespread technology which consumes large amounts of energy.  Triplet-triplet annihilation photochemical upconversion is a method of converting light to a higher frequency.  Here, we show theoretically that photochemical upconversion can be applied to Watt-scale lighting, with performance closely approaching the 50\% quantum yield upper limit.  We describe the dynamic equilibrium of an efficient device consisting of an LED, an upconverting material, and an optical cavity from optical and thermal perspectives.
\end{abstract}

\pacs{}

\maketitle

\section{Potential Advantages of a Photochemical Upconversion Light Emitting Diode}
Triplet-triplet annihilation photochemical upconversion is a technique for converting light to a higher frequency.  It can be relatively efficient at low intensities because it can be exothermic.\cite{cheng2010kinetic,tayebjee2015beyond,dilbeck2017harnessing,hill2016integrated,hill2015photon}  Photochemical upconversion has been demonstrated using abundant chemical elements and solution-phase synthesis.\cite{gray2017loss,peng2016developing}  Therefore it has the potential to be a cheap, widespread technology.

Red light emitting diodes (LEDs) are cheaper, more efficient, and longer lasting than blue LEDs.\cite{held2016introduction,muramoto2014development}  Therefore, it may be cheaper to combine red LEDs and upconversion technology to make blue, ultraviolet, or white LEDs.  While there are many reports discussing the application of upconversion to solar energy capture,\cite{frazer2017optimizing,zeng2017molecular,pedrini2017recent} the first report of a lighting application only appeared recently.\cite{von2018add}  Lighting is important for safety\cite{wanvik2009effects,gaston2014benefits} and consumes 31\,GW in the United States alone.\cite{eia2017}

Here, we quantitatively describe feasible conditions under which the upconversion light source will operate near its theoretical quantum yield upper limit, which is 50\%.  The long-term reliability of photochemical upconversion\cite{zhou2018examining} is an open question which is beyond the scope of this work.  We argue that if future work demonstrates reliability against the degradation mechanisms, which are oxygen permeation and photochemical degradation, upconversion LEDs will be a superior lighting solution.
\section{Upconversion LED Architecture}
\begin{figure*}
 \includegraphics[width=1\textwidth]{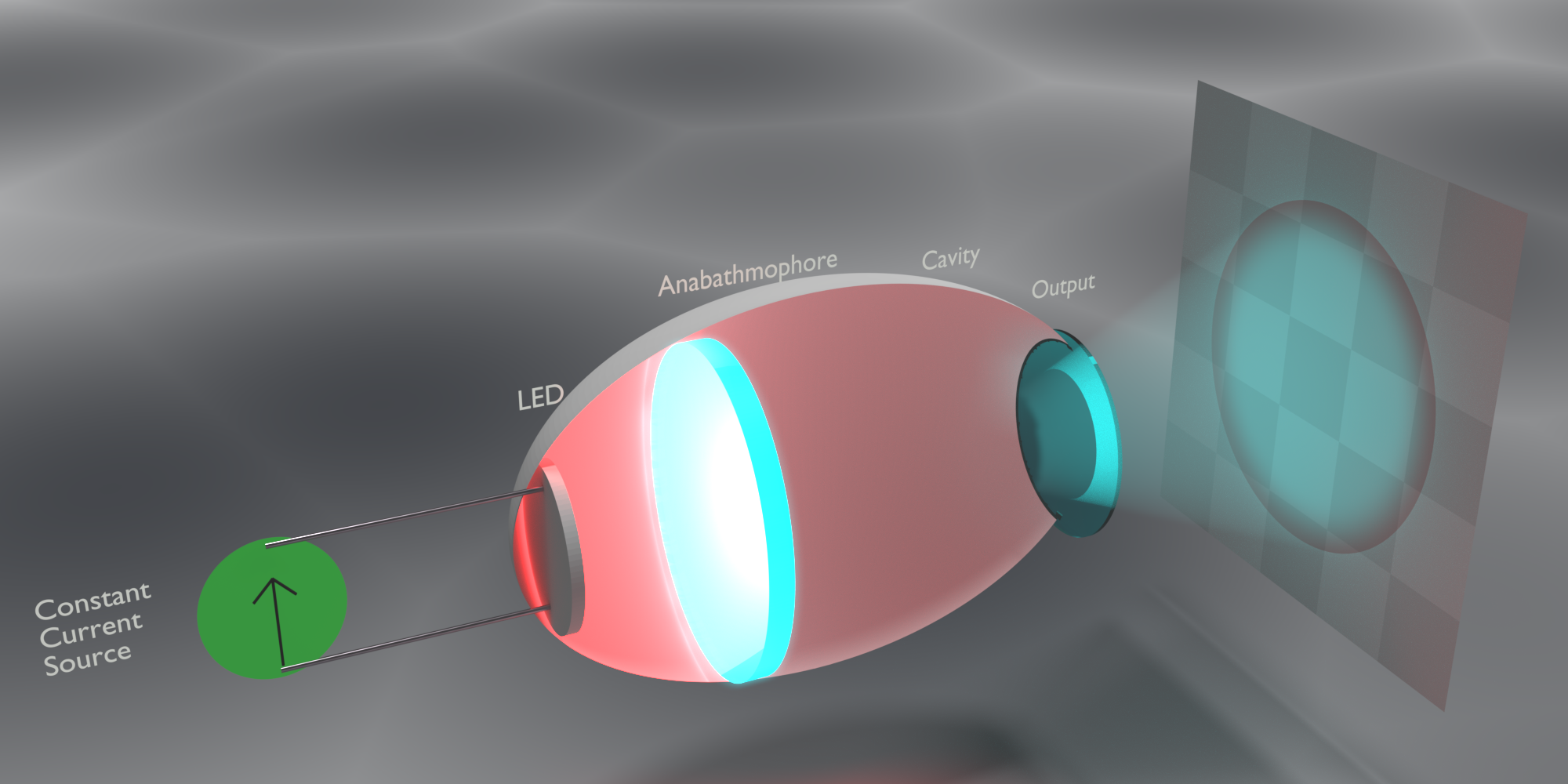}
 \caption{Cartoon of an upconversion LED.  The device consists of a constant current source, LED, cavity, anabathmophore for light conversion, and an output coupler.  Not shown:  heat sink and LED spectral filter.\label{fig:ucled}}
 \end{figure*}

The upconversion LED consists of three parts.  They are illustrated in Figure \ref{fig:ucled}.  The LED, which converts current to light, is the first part.  The anabathmophore, which is the component that converts the light to a higher frequency, is second.  An anabathmophore is a special case of a fluorophore where the emission energy level is above the excitation energy.  The name is derived from the Greek words \begin{otherlanguage}{greek}ἀνα \end{otherlanguage}  `up', \begin{otherlanguage}{greek}βαινω \end{otherlanguage} `to go' and \begin{otherlanguage}{greek}φέρω \end{otherlanguage} `to bring'.  Finally, the optical cavity ensures the LED and anabathmophore are efficiently linked, but the upconversion escapes.
\subsection{LED}
The LED is a junction of a p-doped semiconductor with an n-doped semiconductor.\cite{held2016introduction}  When current flows through the LED, electrons and holes annihilate at the junction, producing light.  LEDs have recently become widespread lighting products.\cite{pust2015revolution}  Despite their high fabrication costs, long lifespans and high conversion efficiencies make them cheaper than traditional lighting technologies, such as incandescent and fluorescent lights.  LEDs are available with high brightnesses and relatively narrow spectra compared to incandescent lights, two features which are necessary for efficient photochemical upconversion LEDs.  In the discussion below, a laser or VCSEL\cite{koyama2006recent} could be substituted for an LED, with no change to the results.  Here, we model the efficiency and heat output at the component level of detail.
\subsection{Photochemical Upconversion}
\begin{figure*}
\includegraphics[width=\textwidth]{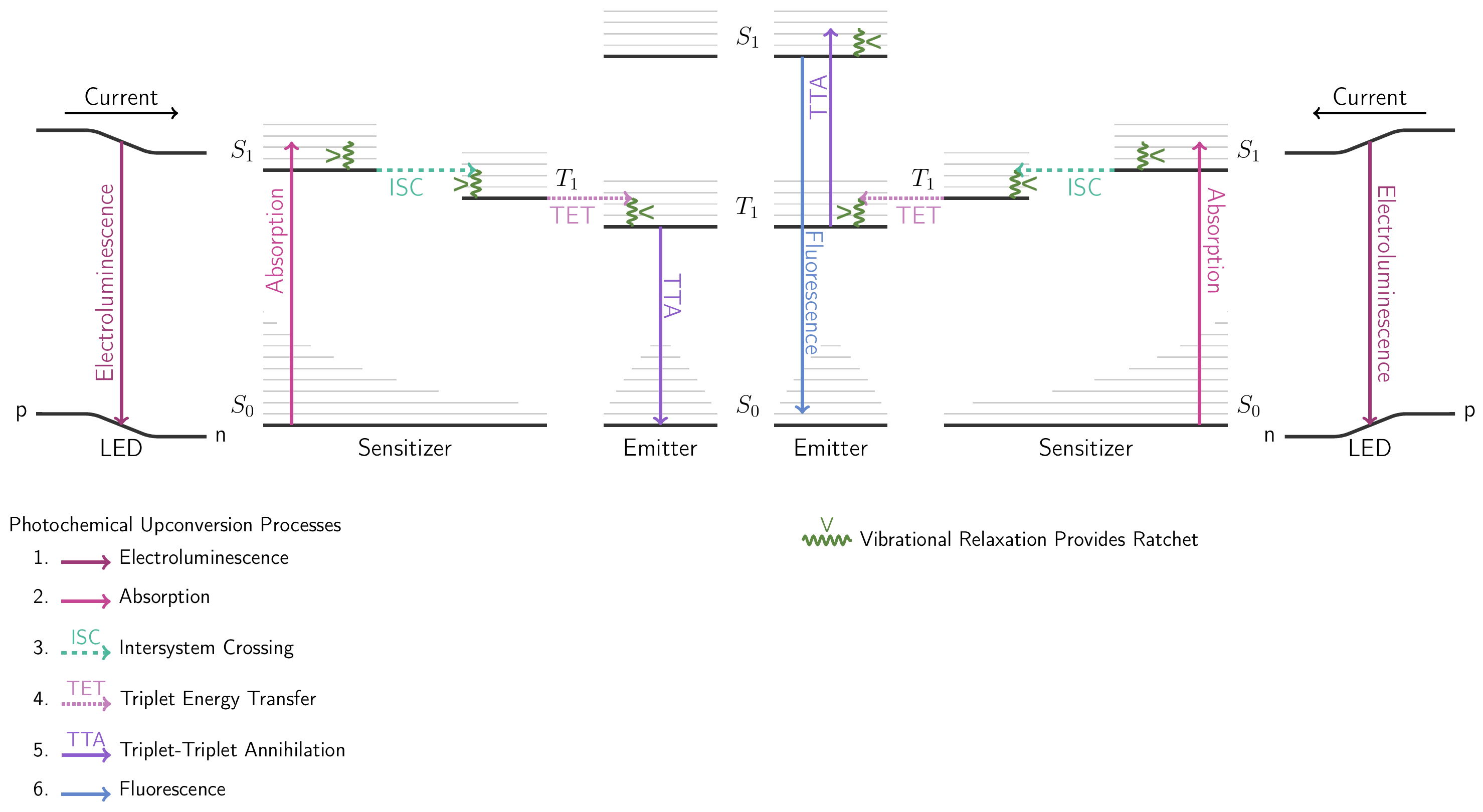}
\caption{Energy level diagram for the six physical processes in a photochemical upconversion LED.  $S_n$ indicates the $n$th singlet state.  $T_1$ indicates the first triplet excited state.  Each process is exothermic, ensuring the device efficiently ratchets its way to a higher energy.  Not to scale.\label{fig:ucdiagram}}
\end{figure*}

Photochemical upconversion is a five-step process which occurs in bimolecular systems.  It is illustrated in steps two to six of Energy Level Diagram \ref{fig:ucdiagram}.  The two molecules are a sensitizer and an emitter, of which there are numerous examples.\cite{peng2016developing,guo2012room,cui2013zinc,islangulov2005low,wu2011organic,yu2015triplet,penconi2013new,deng2015photochemical,wu2015solid,Baluschev_2007,cheng2010efficiency}  The sensitizer absorbs the LED electroluminescence.  This creates a singlet exciton, which converts to a more stable triplet exciton as the sensitizer undergoes intersystem crossing.  Next the sensitizer transfers the triplet exciton to an emitter.  Two emitter triplet excitons undergo Auger exciton-exciton annihilation,\cite{o1999auger,laszlo2013unexpectedly} producing one singlet exciton with a higher energy.  Finally, the singlet exciton undergoes fluorescence.  

The anabathmophore performs spontaneous conversion of light to a higher frequency, which is counter-intuitive because normal fluorophores convert light to a lower frequency.  Photochemical upconversion has a maximum quantum yield of \num{.5} because two quanta are converted to one.  However, the energy efficiency can locally be greater than one, if the emitter is selected to be endothermic and entropy increases.\cite{cheng2011entropically}  In practice, energy levels are selected to make upconversion exothermic in order to prevent reverse operation.\cite{dover2018endothermic}  In this work, we model the anabathmophore rate equations, optical coupling, and component-level heat dissipation.

\subsection{Cavity}
Recently, an LED-driven anabathmophore was demonstrated.\cite{von2018add}  However, the LED emission was not effectively concentrated in the anabathmophore.  Since upconversion relies on a high density of excitons, optical concentration is important to energy efficiency.  An optical cavity can reflect the LED light to achieve a high excitation rate.   The high excitation rate leads to more rapid exciton annihilation.  Here, we use a simplistic model of the cavity at the surface reflectivity level of detail.

The anabathmophore in a cavity resembles an optically excited laser.  However, it does not need to be a laser.  Its operating principle does not rely on stimulated emission.  Upconversion does not have a power threshold which is required for the device to turn on.  A lasing anabathmophore can be distinguished from a regular anabathmophore by dramatically reduced beam divergence.

\section{Upconversion Efficiency}
The quantum yield of photochemical upconversion $\Phi_{\text{UC}}$ is conventionally described in terms of four of the processes which take place within the anabathmophore:\cite{cheng2010efficiency}
\begin{align}
	\Phi_{\text{UC}}&=\frac12\Phi_{\text{ISC}}\Phi_{\text{TT}}\Phi_{\text{TTA}}\Phi_{\text{F}}\label{eq:phiuc}
\end{align}
The symbols are listed in Table \ref{tab:symbols}.
Near-perfect quantum yield is routinely achieved for intersystem crossing in the sensitizer ($\Phi_{\text{ISC}}$) and triplet exciton transfer from the sensitizer to the emitter ($\Phi_{\text{TT}}$).  The quantum yield of triplet-triplet annihilation $\Phi_{\text{TTA}}$ is sensitive to the properties of the LED and will be discussed in depth.  Finally, the fluorescence quantum yield $\Phi_{\text{F}}$ of the emitter in the singlet excited state can be nearly perfect.  In the special case where the emitter possesses a second triplet excited state just above the energy of the singlet exciton, $\Phi_\text{F}$ may be reduced by thermally activated intersystem crossing.\cite{ware1965effect,bennett1966radiationless,lim1966temperature,gillispie1976t}  For emitters with this property, the thermal coupling of the anabathmophore to the LED will have greater significance to $\Phi_\text{F}$.

 \begin{table*}
 \caption{List of symbols with values used in calculations\label{tab:symbols}}
 \begin{ruledtabular}
	 \renewcommand{\arraystretch}{.3}%
 \begin{tabular}{llrr}
	 Symbol&Description&Value&Reference\\
	 $\Phi_{\text{ISC}}$&Intersystem crossing quantum yield&1&\cite{wang2004photophysical,perun2008singlet}\\
	 $\Phi_{\text{TT}}$&Triplet transfer quantum yield&1&\cite{gray2017loss}\\
	 $\Phi_\text{F}$&Fluorescence quantum yield&1&\\
	 $\eta_c$&Singlet yield of exciton annihilation&1&\cite{hoseinkhani2015achieving}\\
	 $f_2$&Proportion of triplet excitons annihilated&&\cite{schmidt2014photochemical}\\
	 $k_1$&Triplet exciton decay rate\footnote{Conventionally reported at a reference temperature\label{fn1}}&\num{e4}\,\si{\per\s}&\cite{cheng2010kinetic,gray2017loss}\\
	 $k_2$&Triplet annihilation rate\footnotemark[1]&\num{e-12}\,\si{\cm^3\per\s}&\cite{frazer2017optimizing}\\
	 $[S]$&Sensitizer concentration&\num{e-3}\,\textsc{m}&\cite{gholizadeh2018photochemical}\\
	 $k_B$&Boltzmann Constant&&\\
	 $E_v$&Energy of vibrational quantum&0.37\,eV&\cite{siebrand1967radiationless}\\
	 $E_A$&Activation energy of triplet annihilation&-1\,eV&\\
	 $P_d$&LED power dissipation&6.6\,W\footnote{LED ENGIN LZ4-00R208 specification sheet, 2018}\\
	 $\phi_r$&LED radiant flux&3\,W\footnotemark[2]&\\
	 $E_d$&Energy per photon from red LED&1.88\,eV\footnotemark[2]&\\
	 $E_U$&Energy per photon from cyan emitter&2.5\,eV&\\
	 $N_s$&Number of sensitizer molecules&\num{e-10}\,\si{\mol}&\\
	 $\epsilon$&Sensitizer molar absorptivity&58,000\,M$^{-1}$cm$^{-1}$&\cite{frazer2017optimizing}\\
	 $x$&Anabathmophore thickness&$10^{-5}$\,m&\\
	 &Anabathmophore area&$10^{-5}$\,m$^2$&\\
	 $A_L$&LED surface area in cavity&$10^{-6}$\,m$^2$&\\
	 $A_c$&Reflector surface area in cavity&$10^{-3}$\,m$^2$&\\
	 $\alpha_L$&LED absorptance&0.5&\\
	 $\alpha_c$&Reflector absorptance&0.01&\\
	 $E_{0r}$&LED Radiant efficacy\footnotemark[1]&.44\footnotemark[2]&\\
	 $k_e$&LED droop&-0.005&\cite{hui2009general}\\
	 $k_h$&LED heating&.56\footnotemark[2]&\cite{qin2009simple}\\
	 $T_0$&Reference temperature&300\,K\footnotemark[2]&\\
	 $T_a$&Ambient temperature&300\,K&\\
	 $R_{jc}$&LED junction-case thermal resistance&2.8\,\si{\K\per\W}\footnotemark[2]&\\
	 $R_{hs}$&Heat sink thermal resistance&1\,\si{\K\per\W}&\\
	 $R_{Uc}$&Anabathmophore-case thermal resistance&10\,\si{\K\per\W}&\\
	 $R_{Uj}$&Anabathmophore-LED junction thermal resistance&no effect&\\
	 \\Results\\\hline
	 $\Phi_{\text{UC}}$&Upconversion quantum yield&.42&\cite{schmidt2014photochemical}\\
	 $\Phi_{\text{TTA}}$&Annihilation quantum yield&.84&\cite{macqueen2014action}\\
	$[^3A^*]$&Triplet exciton concentration&\num{1.0}\,mM&\\
	 $k_\phi$&Excitation rate&\num{7.8e28}\,\si{\per\mol \per\second}&\\
	 $T_U$&Anabathmophore temperature&315\,K&\\
	 $P_{UC}$&Anabathmophore heat output&1.1\,W&\\
	 $\Phi_{cL}$&Yield of LED emission captured by sensitizer&.85&\\
	 $s$&Fraction of energy lost to heat during upconversion&.34&\\
 \end{tabular}
 \end{ruledtabular}
 \end{table*}

\subsection{Steady-State Annihilation Quantum Yield}
We will now examine the annihilation yield to uncover the optical coupling of an anabathmophore to an LED.
\begin{align}
	\Phi_{\text{TTA}}&=\eta_cf_2\label{eq:phitta}
\end{align}
where $\eta_c$ is the proportion of annihilation events which produce a singlet exciton in an emitter molecule.  This parameter captures the density of states and matrix elements for annihilation processes.  It seems reasonable to assume that $\eta_c$ is insensitive to heating by an LED if the heating does not substantially change the alignment of energy levels, including the quintet state and second triplet state,\cite{dick1983accessibility,gray2014triplet,saltiel1981spin,tayebjee2017quintet,cheng2010kinetic} in the emitter molecule.  There are few reports of measurements of $\eta_c$.\cite{hoseinkhani2015achieving,cheng2010kinetic}  

$f_2$ is the proportion of triplet excitons which decay by annihilation.  It is the ratio of the desirable kinetics to the total decay rate:\cite{schmidt2014photochemical}
\begin{align}
	f_2&=\frac{k_2[^3A^*]}{k_2[^3A^*]+k_1}.\label{eq:f2}
\end{align}
$k_1$ is the small decay rate of a triplet exciton in an isolated emitter.  $k_2$ is the second order rate constant for triplet annihilation.  $[^3A^*]$ is the triplet exciton concentration in the emitter molecule.  Efficiency is achieved when $k_2$ and $[^3A^*]$ are large.

The steady state triplet exciton concentration in terms of device parameters is:\cite{schmidt2014photochemical}
\begin{align}
  [^3A^*]&=\frac{-k_1+\sqrt{k_1^2+4k_2k_\phi\Phi_{\text{ISC}}\Phi_{\text{TT}}[S]}}{2k_2}\label{eq:tripletconc}
\end{align}
Where $k_\phi$ is the excitation rate caused by the LED and $[S]$ is the sensitizer concentration.
\cite{calef1983diffusion} To achieve good radiant flux and energy efficiency, the optical system should be arranged to produce a large $k_\phi$.  This will be discussed in Section \ref{sec:cavperform}.  

\subsection{Triplet exciton decay\label{sec:triplet}}
Thermal coupling between the LED and anabathmophore plays a role in the triplet exciton decay.
The unimolecular triplet exciton decay rate $k_1$ has thermal, collisional, and fixed (e.g. phosphorescence\cite{aulin2015photochemical,gray2017loss}) components.  The thermal relaxation component has the form\cite{siebrand1967radiationless,thevenaz2016thermoresponsive} 
\begin{align}
  e^{-\frac{E_v}{k_BT_U}}\label{eq:k1temp}
\end{align}
where the vibrational energy $E_v$ is typically about 3000\,\si{\per\cm},\cite{siebrand1967radiationless} $k_B$ is the Boltzmann constant,  and $T_U$ is the temperature of the anabathmophore.  The collisional component includes phenomena like triplet quenching by oxygen\cite{schweitzer2003physical} and  by the sensitizer.\cite{gholizadeh2018photochemical}

The triplet exciton decay rate may be increasing, decreasing, or static as a function of temperature depending on the mechanism.  To achieve an energy efficient device, the LED must provide sufficient illumination that $k_\phi[S]$ overwhelms $k_1$.  If this is achieved, the thermal effects of the LED on $k_1$ will become unimportant.  For modelling purposes, we use the thermal relaxation behavior (\ref{eq:k1temp}).

\subsection{Annihilation rate}
The annihilation reaction is a second order reaction.  It is diffusive and has an activation energy.  
Neglecting the range of the reaction and the mobility of the annihilators,
\begin{align}
	k_2&\propto k_BT_U e^{-\frac{E_A}{k_BT_U}}\label{eq:k2temp}
\end{align}
where $E_A$ is the annihilation activation energy and $T_U$ is the temperature of the anabathmophore.  The energy $E_A$ is the potential energy difference between the energy barrier to triplet annihilation and the energy of two emitter triplets.  The barrier cannot be less than the emitter singlet energy.  Ideally, $E_A$ will be negative. The temperature dependence of several aspects of upconversion have been demonstrated.\cite{singh2009influence,thevenaz2016thermoresponsive,askes2017temperature,massaro2016thermally,xu2018ratiometric}  The temperature/$k_2$ relationship (\ref{eq:k2temp}) is ubiquitous in chemistry but, to our knowledge, has not been directly tested with photochemical upconversion.

\begin{figure}
 \includegraphics[width=.5\textwidth]{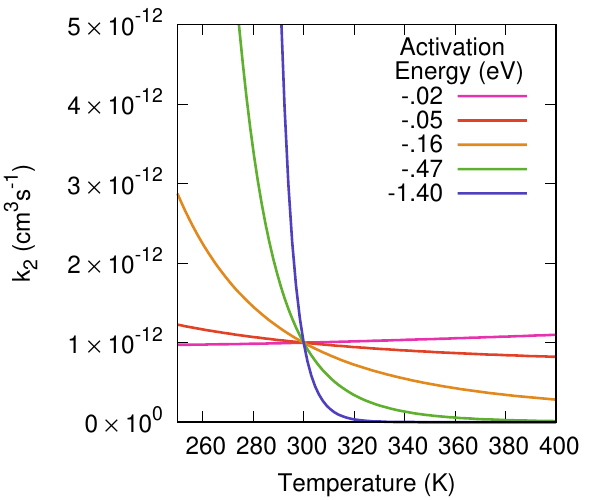}
 \caption{Predicted temperature dependence of the annihilation rate $k_2$ for various activation energies, using 300\,K as a reference point.  If the measured value of $k_2$ is $10^{-12}$\,cm$^3$s$^{-1}$ at $300$\,K, the value of $k_2$ at other temperatures can be predicted based on the activation energy.  To achieve efficiency, upconversion should be slightly exothermic.\label{fig:k2t}}
 \end{figure}

$k_2$ is an important parameter to maximize in order to achieve overall efficiency.  There are three regimes to consider:  First, $0< E_A$ is endothermic upconversion.   This case should only be selected if the upconversion must increase the photon energy by more than a factor of two, as it is inefficient.  For endothermic upconversion, higher temperature operation is always more efficient.  Second, if $0>E_A>-k_BT$, then $k_2$ increases with temperature.  In the high temperature limit, the increase is linear, as shown by the fuchsia curve in Figure \ref{fig:k2t}.  Third, the typical situation is that $0>-k_BT>E_A$ and $k_2$ decreases with temperature.  As shown by the blue curve in Figure \ref{fig:k2t}, the decrease can be rapid.  

If the goal is to produce a white light by combining the LED and upconversion emission, then an activation energy of about -1.4 eV is desired.    The activation energy is determined by the energy of complementary colors in the additive color model, energy losses involved in sensitizer absorption, intersystem crossing, triplet transfer, and singlet fluorescence.  If there is an energy barrier to annihilation which lies above the singlet state, it must also be accounted for in the activation energy of a white LED.

\section{Cavity Performance\label{sec:cavperform}}
As seen in Equation (\ref{eq:tripletconc}), the sensitizer excitation rate $k_\phi$ is a controllable parameter which contributes to the efficiency of photochemical upconversion.  If the radiant flux of the LED is $\phi_r$, then the upper limit on $k_\phi$ is 
\begin{align}
	k_\phi&\le \frac{\phi_r}{E_dN_s}\label{eq:exlimit}
\end{align}
where $E_d$ is the energy per photon emitted by the diode\footnote{We omit the shape and temperature dependence of the LED spectrum throughout.} and $N_s$ is the number of sensitizer molecules.  This shows that the number of sensitizer molecules should be as small as possible to increase the excitation rate $k_\phi$.   Equation (\ref{eq:tripletconc}) indicates the sensitizer concentration should be high.  Considering Equations (\ref{eq:tripletconc}) and (\ref{eq:exlimit}) together, the volume of the anabathmophore should be low.  Unlike sunlight,\cite{xie2011concentrated,kalogirou2004solar,badescu1995radius} LED emission can be easily placed in a cavity, making it possible to achieve both a small volume anabathmophore and a high excitation rate.

The cavity will have two loss mechanisms:  the loss occurring from imperfect cavity walls and the loss from absorption of light reflected back to the LED.  The volume of the anabathmophore should be large enough that it absorbs more of the light than is lost to either mechanism.  

Neglecting geometric details, the proportion of LED light injected into a cavity which is absorbed in the sensitizer, $\Phi_{cL}$, is the product of the anabathmophore absorption probability and the cavity reflection probability summed over cavity traverses performed by the LED light:
\begin{widetext}
\begin{align}
	\Phi_{cL}&=\left(1-e^{-\epsilon [S]x}\right)\sum_{n=0}^{\infty}\left( e^{-\epsilon [S]x} \right)^n\left(1-\frac{A_L\alpha_L+A_c\alpha_c}{A_L+A_c}\right)^n,
\end{align}
\end{widetext}
where $\epsilon$ is the sensitizer molar absorptivity used in the Beer-Lambert Law, $x$ is the average thickness of the upconversion material traversed by light crossing the cavity, $A_L$ is the LED area, $A_c$ is the cavity reflector area, $\alpha_L$ is the LED absorptance, and $\alpha_c$ is the cavity reflector absorptance.  The geometric series simplifies to
\begin{align}
	\Phi_{cL}&=
\frac{
\left(1-e^{-\epsilon [S]x}\right)
}{1-
\left(e^{-\epsilon [S]x}\right)
\left(1-\frac{A_L\alpha_L+A_c\alpha_c}{A_L+A_c}\right)}
\label{eq:phicavity},
\end{align}
which is the yield of the coupling between the LED and the anabathmophore.
The resulting excitation rate is 
\begin{align}
	k_\phi&= \frac{\Phi_{cL}\phi_r}{E_dN_s}.\label{eq:kphi}
\end{align}
A higher excitation rate is better.  The excitation rate can be increased by using a high LED brightness, a highly reflective cavity, and a small anabathmophore.
This model is reasonable for a planar geometry with a monochromatic LED, specular reflections, and homogeneous materials.  In other cases, the cavity can be modeled in detail with Monte Carlo methods.

Since no stimulated emission is required, but low cost is often desirable, we pick a Teflon (polytetrafluoroethylene) cavity as a concrete example.  Teflon has absorptance $\alpha_c\approx0.01$.  The diffuse reflectivity of a Teflon cavity is acceptable as sensitization is insensitive to the angle of incidence.  For our model, we assume the LED has $\alpha_c\approx 0.5$ and is about $10^{-6}$\,m$^2$ in area.  We assume the cavity has an internal surface area of $10^{-4}$\,m$^2$, so losses from the LED absorption are not very important.  Since the upconversion must be coupled out of the device, $10^{-5}$\,m$^2$ of the cavity Teflon should be replaced with a relatively expensive dielectric short-wavelength-transmitting filter with $\alpha_c=0.01$ or better.  Such a Bragg filter would transmit the upconverted light and specularly reflect the LED emission back into the cavity.

Owing to the Beer-Lambert law, to optimize $k_\phi$ the upconversion portion of the device should be a thin film.\cite{frazer2017optimizing}  To continue the example, the sensitizer concentration might be around $1$\,mM with anabathmophore thickness \num{e-5}\,\si{\meter}, area \num{e-5}\,\si{\meter^2} and \num{e-10}\,\si{\mole} of sensitizer.

The LED can be covered by a filter\cite{von2018add} which reflects the upconverted light, eliminating one source of self-absorption.  If the filter is a well-designed dielectric thin film, no significant losses of the LED light will occur.  A good anabathmophore design will possess a low molar absorptivity at the emission wavelength.  Our model assumes self-absorption of the upconverted light is negligible because the anabathmophore is thin.
\section{Steady-State Thermal Performance}
The radiant flux of an LED, $\phi_{r}$, is to be distinguished from the human vision-weighted luminous flux typically specified by visible LED manufacturers; upconversion can be pumped by infrared LEDs.\cite{deng2013near,zou2012broadband,wu2016solid,fuckel2011singlet,yakutkin2008towards,singh2010supermolecular,huang2015hybrid}   As LED physics can vary considerably between designs and operating regimes, we draw from a component-level phenomenological theory\cite{hui2009general} which is valid over a limited range and does not include any wavelength conversion device.  $\phi_r$ is temperature sensitive:\cite{hui2009general}
\begin{widetext}
\begin{align}
	\phi_{r}(P_d)&=E_{0r}\left\{ \left[ 1+k_e\left( T_a-T_0 \right)\right] P_d+k_ek_h\left( R_{jc}+R_{hs} \right)P_d^2 \right\}\label{eq:ledout}
\end{align}
\end{widetext}
$E_{0r}$ is the radiant efficacy specification at a reference temperature $T_0$. $k_e$, which is negative, models the LED droop.  Droop is the underperformance of LEDs at high current, which has several causes, including annihilation events.\cite{iveland2013direct,pozina2015dislocation,kioupakis2011indirect,binder2013identification,verzellesi2013efficiency,cho2013efficiency,karpov2015abc}  $T_a$ is the ambient temperature.  $P_d$ is the diode input power, as measured from the current and voltage.  $k_h$ is the proportion of the power $P_d$ which is dissipated as heat.  $R_{jc}$ is the thermal resistance\cite{luo2016heat} between the LED junction and the LED case/heat sink contact.  $R_{hs}$ is the thermal resistance of the heat sink.
 \begin{figure}
 \includegraphics[width=.2\textwidth]{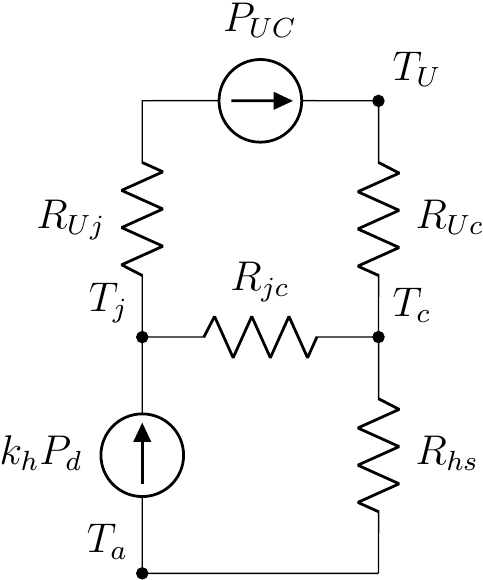}
 \caption{Thermal circuit diagram of the temperatures, heat loads, and thermal resistances in an upconversion LED system.\label{fig:thermal}  $T_a$ is the ambient temperature.  $T_j$ is the LED junction temperature.  $T_U$ is the anabathmophore temperature.  $T_c$ is the case temperature.  The heat loads are from the LED ($k_hP_d$) and anabathmophore ($P_{UC}$).}
 \end{figure}

The thermal circuit, expanded to include the heat load of the anabathmophore, is illustrated in Figure \ref{fig:thermal}.
The anabathmophore is in thermal contact with the case (thermal resistance $R_{Uc}$) and  the LED (thermal resistance $R_{Uj}$).  The heat output of the anabathmophore and cavity is
\begin{widetext}
\begin{align}
	P_{UC}&=\phi_r\left( (1-\Phi_{cL}) + \Phi_{cL}(1-2\Phi_{\text{UC}}) + \Phi_{cL}\Phi_{\text{UC}}s \right)\label{eq:ucheat}.
\end{align}
\end{widetext}
The first term is the cavity loss because the cavity coupling $\Phi_{cL}$ is less than one.
The second term is the heat output owing to triplets which decay without becoming singlets.  This heat is computed from the upconversion quantum yield $\Phi_{UC}$.  
The third term is the heat output from exothermic upconversion.  
The fraction of energy which is lost to heat  during upconversion, $s$, can be computed from the area-normalized LED emission spectrum as a function of energy $\psi_L(E)$ and the anabathmophore emission spectrum $\psi_{UC}(E)$:
\begin{align}
	s&=\frac{\int 2E\psi_{L}(E)-E\psi_{UC}(E)dE}{2\int E\psi_{L}(E)dE}.
\end{align}
The losses which contribute to $s$ include cumulative donor and acceptor Stokes, intersystem crossing, transfer, and annihilation energy shifts.
We approximate $s$ by assuming the spectra are monochromatic.  The way the upconversion quantum yield $\Phi_{\text{UC}}$ is included in the heat output in Model (\ref{eq:ucheat}) neglects any phosphorescence which may, inadvertently or to generate another color, escape from the device.  An efficient anabathmophore produces negligible phosphorescence.\cite{aulin2015photochemical,gray2017loss}

The temperatures of the combined devices can be determined using Kirchhoff nodal analysis.  The diode temperature is 
\begin{align}
	T_a+R_{hs}k_hP_d+R_{jc}\left( k_hP_d-P_{UC} \right)\label{eq:diodetemp}
\end{align}
and the anabathmophore temperature is 
\begin{align}
	T_U&=T_a+R_{hs}k_hP_d+R_{Uc}P_{UC}\label{eq:tu}.
\end{align}
In Section \ref{sec:triplet}, we explained that the anabathmophore temperature alters the efficiency of upconversion by changing the annihilation rate constant.  Here, we find that the heat from the LED decreases the efficiency of the upconversion.

The LED performance in Equation (\ref{eq:ledout}) can be revised to include the anabathmophore's contribution to the thermal circuit:
\begin{widetext}
\begin{align}
	\phi_{r}(P_d)&=E_{0r}\left\{ 1+k_e\left[ T_a-T_0 +R_{jc}\left( k_hP_{d}-P_{UC} \right)+R_{hs}k_hP_d \right]
	\right\} P_d. \label{eq:ledout2}
\end{align}
\end{widetext}
The heat output from the anabathmophore $P_{UC}$ increases the LED radiant flux $\phi_r$ because heat is directed to the LED case instead of the junction.

\section{Computational Methods}
We numerically solved the system of simultaneous equations (\ref{eq:phiuc}), (\ref{eq:phitta}), (\ref{eq:f2}),  (\ref{eq:tripletconc}),  (\ref{eq:kphi}), (\ref{eq:ucheat}), (\ref{eq:tu}), and  (\ref{eq:ledout2}) with $k_1$ and $k_2$ scaled with anabathmophore temperature $T_U$ according to expressions (\ref{eq:k1temp}) and  (\ref{eq:k2temp}) respectively.  The values of the physical parameters we used are listed in Table \ref{tab:symbols}.  In the next section, some of these parameters are varied one at a time to show their importance to device performance.  The final result, the upconversion LED radiant flux, is $\Phi_{\text{UC}}\Phi_{cL}\phi_r(E_U/E_d)$.  That is the product of the upconversion quantum yield, the cavity coupling yield, the LED radiant flux, and the gain caused by the spectral shift from the LED emission energy $E_d$ to the upconversion emission energy $E_U$ of the emitter.
\section{Results\label{sec:compres}}
The solution to the model is listed in Table \ref{tab:symbols}.  The device wall-plug efficiency is 20\% 
and the upconversion radiant flux is 1\,W.  High efficiency is achieved because the triplet exciton concentration is high.  The concentration of the emitter, which played no role in the calculation, must be higher than the triplet exciton concentration in order for the results to be physically correct.  

The 20\% wall-plug efficiency is lower than the approximately 30\% efficiency of commercial blue LEDs, or the 80\% achieved in laboratory devices.\cite{kuritzky2017high}
In the model, the most important factor reducing the system efficiency is the red LED's efficiency.  Next most important is the fundamental upper limit that photochemical upconversion cannot exceed: 50\% quantum yield.  The spectral shift of the upconversion can help make up for this limit.  In the model we selected complementary colors suitable for generating white light, which reduces the spectral shift.  We chose an increase in the energy per photon of just 33\%.  Finally, the coupling of the LED to the anabathmophore is the most efficient component, achieving 85\% with our absorptance assumptions.

\subsection{Triplet exciton lifetime}
\begin{figure}
 \includegraphics[width=.5\textwidth]{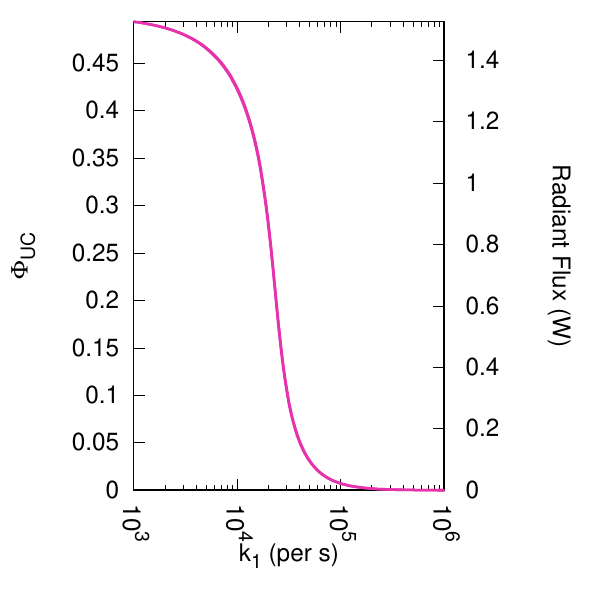}
 \caption{Calculated upconversion quantum yield $\Phi_{\text{UC}}$ and upconversion output power as a function of the emitter molecule triplet exciton decay rate $k_1$.  The device efficiency improves as $k_1$ gets smaller.  Readily achievable decay rates push the system close to the quantum yield upper limit, $\Phi_{\text{UC}}\le 0.5$.  All other parameters are as specified in Table \ref{tab:symbols}. \label{fig:k1}}
 \end{figure}

The annihilation yield depends on the balance of the different triplet exciton decay rates.  In Figure \ref{fig:k1}, we show that, under the conditions listed in Table \ref{tab:symbols}, the triplet exciton concentration is so high that triplet exciton utilization is nearly perfect when the unimolecular triplet exciton decay $k_1$ is less than \num{e4} \si{\per\second}.  With rubrene as an emitter, \num{8e3} \si{\per\second} has been achieved,\cite{cheng2010kinetic} while for 9-(4-phenylethynyl)-10-phenylanthracene, $k_1$ is around \num{5e2} \si{\per\second}.\cite{gray2017loss}  Since $\Phi_{\text{TTA}}$ must be no greater than \num{.5} owing to energy conservation, further improvements in $k_1$ are not necessary.  However, the introduction of molecular oxygen to the device must be prevented because oxygen is an efficient triplet quencher.\cite{schweitzer2003physical}

 \subsection{Yields}
\begin{figure}
 \includegraphics[width=.5\textwidth]{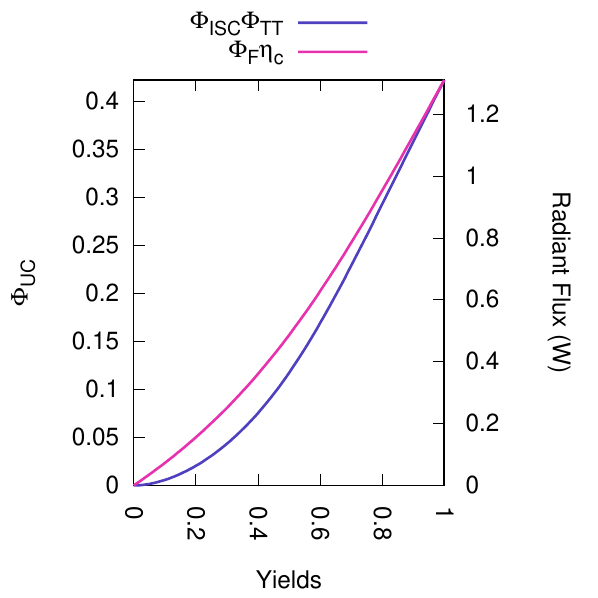}
 \caption{Calculated upconversion quantum yield $\Phi_{\text{UC}}$ and upconversion output power as a function of the yields $\Phi_{\text{ISC}}\Phi_{\text{TT}}$ (blue) and $\Phi_{\text{F}}\eta_c$ (red).  Reduced $\Phi_{\text{ISC}}$ or $\Phi_{\text{TT}}$ decrease device efficiency before the annihilation step.  While they are physically different, those two yields play the same mathematical role in the model, so they are presented as a product.  Reduced $\Phi_{\text{F}}$ or $\eta_c$ decrease device efficiency after the annihilation step, so they are slightly less important.   This pair of yields only appear in the calculation as a product, so they are presented as a product. All other parameters are as specified in Table \ref{tab:symbols}. \label{fig:yields}}
 \end{figure}

 Based on experimental results, the yields $\Phi_{\text{ISC}}$, $\Phi_{\text{TT}}$, $\Phi_{\text{F}}$, and $\eta_c$ may be perfect.  If they are not, device performance is reduced.  The first two, the yields of intersystem crossing and triplet transfer, are more important because they play a role in causing annihilation, which is a nonlinear process.  Figure \ref{fig:yields} shows that devices with impaired $\Phi_\text{F}\eta_c$ outperform devices with impaired $\Phi_{\text{ISC}}\Phi_{\text{TT}}$ because the fluorescence yield and singlet yield of exciton annihilation are not involved in that nonlinearity.  The rate constants are presented multiplied together where their results are not distinguishable.

 \subsection{Sensitizer concentration}
\begin{figure}
 \includegraphics[width=.5\textwidth]{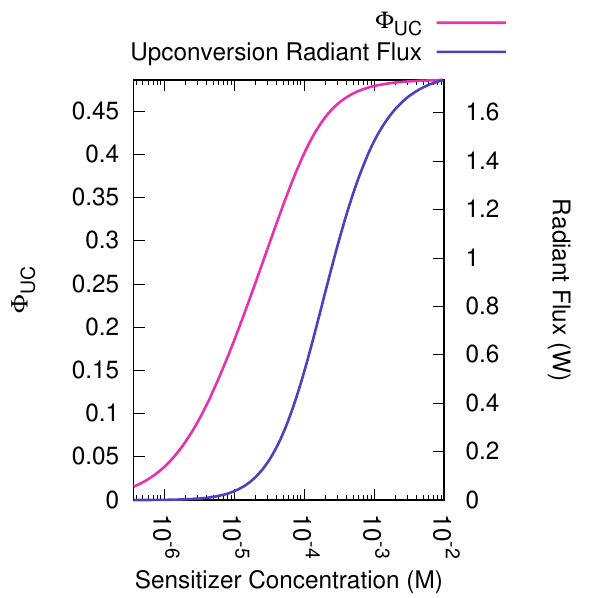}
 \caption{Calculated upconversion quantum yield $\Phi_{\text{UC}}$ (red) and upconversion output power (blue) as a function of sensitizer concentration.  Higher sensitizer concentration is better.  All other parameters are as specified in Table \ref{tab:symbols}. \label{fig:sensitizer}}
 \end{figure}

 Reduction of the sensitizer concentration is harmful.\cite{frazer2017optimizing}  The sensitizer concentration is critically important because it determines triplet concentration.  The sensitizer concentration needs to be high to enable triplets to encounter partners for annihilation.  Figure \ref{fig:sensitizer} indicates that millimolar concentrations are desirable.   1 mM has been achieved in solution\cite{gholizadeh2018photochemical} and future solid-phase devices should have much higher sensitizer concentrations.

\subsection{Electrical power}
\begin{figure}
 \includegraphics[width=.5\textwidth]{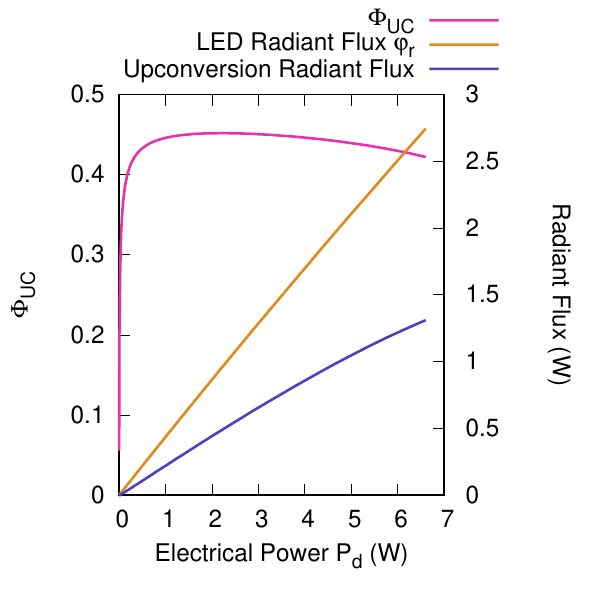}
 \caption{Calculated upconversion quantum yield $\Phi_{\text{UC}}$ and LED/upconversion output power as a function of the electrical power consumed by the device.  Efficient performance is achievable with commercially available high-power LEDs.  All other parameters are as specified in Table \ref{tab:symbols}. \label{fig:power}}
 \end{figure}

Efficient upconversion requires high excitation rates.  In Figure \ref{fig:power}, we show that \si{Watt}-scale LEDs, which are commercially available, can provide sufficient optical power.  Both the LED (orange) and anabathmophore (blue) response to the power consumed are only slightly worse than linear.  At \si{milliWatt}-scale driving power, however, the triplet excitons fail to find partners for annihilation, and the device efficiency plummets.  $\Phi_\text{TTA}$ has a maximum at 2.2\,W.  Above this electrical power, we find detrimental thermal effects on the upconversion rate constants.

\subsection{Cavity performance}
\begin{figure}
 \includegraphics[width=.5\textwidth]{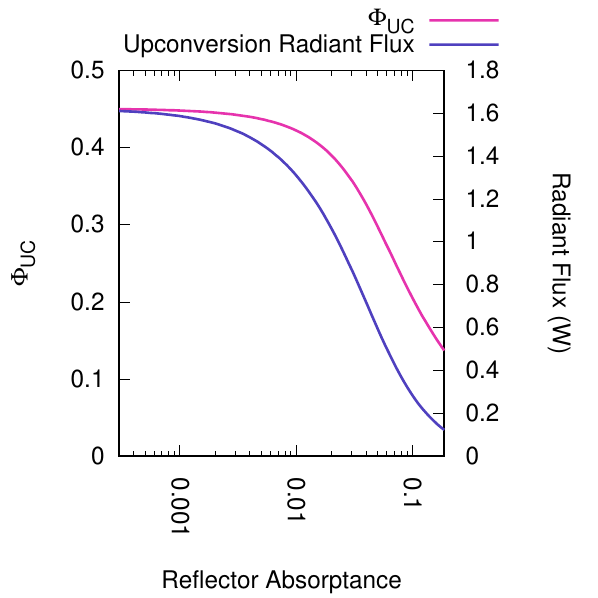}
 \caption{Calculated upconversion quantum yield $\Phi_{\text{UC}}$ and upconversion output power as a function of the cavity reflector absorptance $\alpha_c$.  Low-cost Teflon has an absorptance of about 0.01, which is sufficient for good upconversion quantum yield $\Phi_{\text{UC}}$.  However, even lower absorptance produces a greater power output owing to a higher cavity efficiency $\Phi_{cL}$.  All other parameters are as specified in Table \ref{tab:symbols}. \label{fig:absorptance}}
 \end{figure}

In Figure \ref{fig:absorptance}, we show that an optical cavity is necessary to achieve efficient operation.  If the cavity's reflector has absorptance greater than \num{0.01}, the yield of photochemical upconversion decreases rapidly.  Our simple cavity model assumes multiple, randomized reflections, which are required for efficient operation.  Therefore it is not reliable at very high absorptances.  

 A high upconversion quantum yield $\Phi_{\text{UC}}$ is necessary but not sufficient.  The cavity must also effectively direct the LED electroluminescence to the sensitizer.  Figure \ref{fig:absorptance}  shows that as the absorptance decreases, near \num{0.01} the $\Phi_{\text{UC}}$ is saturating, but the device power output is still improving.  20\% overall efficiency can be achieved with low cost (e.g. Teflon) diffuse reflectors with \num{0.01} absorptance.  Marginally lower absorptance could be achieved using a more costly dielectric reflector.\cite{stanley1994ultrahigh}
 Cavity performance is ultimately limited by the self-absorption of the LED.  While the LED can be prevented from absorbing the upconversion light by coating it with a dielectric filter,\cite{von2018add} it cannot be prevented from absorbing back-reflections of its own electroluminescence without an impractical optical isolator.\cite{schulz1989wavelength,gauthier1986simple}
\subsection{Thermal resistance}
\begin{figure}
 \includegraphics[width=.5\textwidth]{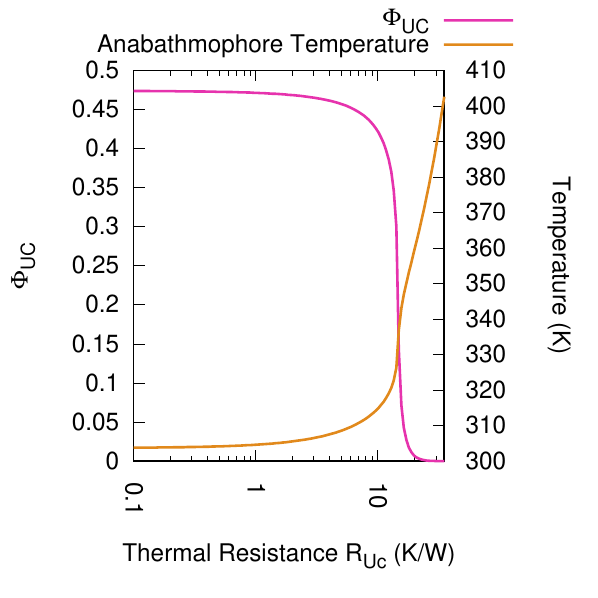}
 \caption{Calculated upconversion quantum yield $\Phi_{\text{UC}}$ and anabathmophore temperature $T_U$ as a function of the anabathmophore-to-case thermal resistance $R_{Uc}$. If the anabathmophore is highly insulated, it gets hot.  The heat leads to unfavorable kinetic parameters $k_1$ and $k_2$, producing declining efficiency.  The rate constants are compromised before the decomposition temperature is reached.  All other parameters are as specified in Table \ref{tab:symbols}. \label{fig:ruc}}
 \end{figure}

Heating is a well-known problem in high-power light-emitting diodes, which are typically designed with heat dissipation in mind.\cite{cheng2012heat,lin2011heat,weng2009advanced,hui2009general,yung2014thermal,yung2014heat} 
Figure \ref{fig:ruc} shows that the upconversion LED must be designed with a reasonable thermal resistance $R_{Uc}$ between the anabathmophore and the case.  If the thermal resistance exceeds \num{10}\,\si{\K\per\W}, then the anabathmophore temperature rises rapidly.  As a result, the rate constants $k_1$ and $k_2$ become less favorable, reducing the upconversion efficiency.  Above 20\,\si{\K\per\W}, nearly all the energy is lost as heat.  Thermal resistances less than 1 \si{\K\per\W} are common in commercial electronic devices.
\subsection{Ambient temperature}
\begin{figure}
 \includegraphics[width=.5\textwidth]{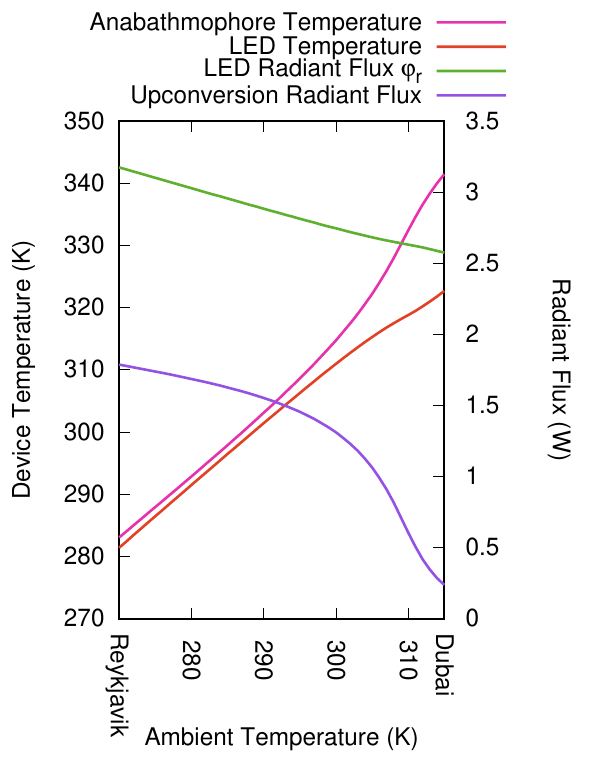}
 \caption{Calculated LED/anabathmophore temperature and LED/upconversion radiant flux as a function of ambient temperature.  The upconversion LED is brighter at lower temperatures.  All other parameters are as specified in Table \ref{tab:symbols}. \label{fig:ambient}}
 \end{figure}

To be practical for outdoor use, light sources need to operate efficiently at a wide range of ambient temperatures $T_a$.  In Figure  \ref{fig:ambient}, we calculate the decline in the brightness of the upconversion LED which is caused by elevated ambient temperatures.  The LED component has a declining output, which is the well-known LED droop.  The upconversion output has a more severe droop, which is caused by the anabathmophore's nonlinear response.  The reduction in the anabathmophore efficiency at high temperatures increases the anabathmophore's heat dissipation, which has Equation (\ref{eq:tu})'s feedback effect on the anabathmophore's temperature.  The increase in the anabathmophore's heat dissipation mitigates the diode temperature increase, as indicated in Expression (\ref{eq:diodetemp}).

 \section{Conclusion}
We show that there are a wide range of conditions under which upconversion LEDs can operate efficiently.  Currently achievable triplet exciton lifetimes are sufficient to avoid losses from unimolecular triplet exciton decay.  Commercially available LEDs have sufficient brightness to drive the device.  Modest performance optical cavities can effectively couple the LED and anabathmophore.  A basic heat sink is required to ensure the anabathmophore does not overheat.  However, at high ambient temperatures, care must be taken to improve the thermal resistance or rate constants to be better than our assumed values.  Otherwise, the device brightness can decline in hot weather.  

Upconversion LEDs incorporating effective optical and thermal designs will readily exceed the efficiency of incandescent lights.  They may never achieve the efficiency of the best laboratory LEDs, but may still be an economically competitive technology. At present, the only barrier to implementation of upconversion LEDs is uncertainty about the long-term reliability of anabathmophores.  This reliability should be tested against blue LEDs.  If anabathmophores are shown to be stable, then upconversion LEDs will be suitable for lighting human vision with white light and for blue or ultraviolet illumination in applications such as lithography,\cite{suzuki2008lithography,harburg2013chip,guijt2008maskless} banknote identification,\cite{muramoto2014development} or photochemistry.\cite{kim2012encapsulated,wu2013observation}

\begin{acknowledgments}
This work was supported by the Australian Research Council Centre of Excellence in Exciton Science (CE170100026).
We thank Dr. Andrew Connor for the word anabathmophore.
\end{acknowledgments}

\bibliography{ucled}

\begin{thebibliography}{91}%
\makeatletter
\providecommand \@ifxundefined [1]{%
 \@ifx{#1\undefined}
}%
\providecommand \@ifnum [1]{%
 \ifnum #1\expandafter \@firstoftwo
 \else \expandafter \@secondoftwo
 \fi
}%
\providecommand \@ifx [1]{%
 \ifx #1\expandafter \@firstoftwo
 \else \expandafter \@secondoftwo
 \fi
}%
\providecommand \natexlab [1]{#1}%
\providecommand \enquote  [1]{``#1''}%
\providecommand \bibnamefont  [1]{#1}%
\providecommand \bibfnamefont [1]{#1}%
\providecommand \citenamefont [1]{#1}%
\providecommand \href@noop [0]{\@secondoftwo}%
\providecommand \href [0]{\begingroup \@sanitize@url \@href}%
\providecommand \@href[1]{\@@startlink{#1}\@@href}%
\providecommand \@@href[1]{\endgroup#1\@@endlink}%
\providecommand \@sanitize@url [0]{\catcode `\\12\catcode `\$12\catcode
  `\&12\catcode `\#12\catcode `\^12\catcode `\_12\catcode `\%12\relax}%
\providecommand \@@startlink[1]{}%
\providecommand \@@endlink[0]{}%
\providecommand \url  [0]{\begingroup\@sanitize@url \@url }%
\providecommand \@url [1]{\endgroup\@href {#1}{\urlprefix }}%
\providecommand \urlprefix  [0]{URL }%
\providecommand \Eprint [0]{\href }%
\providecommand \doibase [0]{http://dx.doi.org/}%
\providecommand \selectlanguage [0]{\@gobble}%
\providecommand \bibinfo  [0]{\@secondoftwo}%
\providecommand \bibfield  [0]{\@secondoftwo}%
\providecommand \translation [1]{[#1]}%
\providecommand \BibitemOpen [0]{}%
\providecommand \bibitemStop [0]{}%
\providecommand \bibitemNoStop [0]{.\EOS\space}%
\providecommand \EOS [0]{\spacefactor3000\relax}%
\providecommand \BibitemShut  [1]{\csname bibitem#1\endcsname}%
\let\auto@bib@innerbib\@empty
\bibitem [{\citenamefont {Cheng}\ \emph
  {et~al.}(2010{\natexlab{a}})\citenamefont {Cheng}, \citenamefont {F\"uckel},
  \citenamefont {Khoury}, \citenamefont {Clady}, \citenamefont {Tayebjee},
  \citenamefont {Ekins-Daukes}, \citenamefont {Crossley},\ and\ \citenamefont
  {Schmidt}}]{cheng2010kinetic}%
  \BibitemOpen
  \bibfield  {author} {\bibinfo {author} {\bibfnamefont {Y.~Y.}\ \bibnamefont
  {Cheng}}, \bibinfo {author} {\bibfnamefont {B.}~\bibnamefont {F\"uckel}},
  \bibinfo {author} {\bibfnamefont {T.}~\bibnamefont {Khoury}}, \bibinfo
  {author} {\bibfnamefont {R.~G.}\ \bibnamefont {Clady}}, \bibinfo {author}
  {\bibfnamefont {M.~J.}\ \bibnamefont {Tayebjee}}, \bibinfo {author}
  {\bibfnamefont {N.}~\bibnamefont {Ekins-Daukes}}, \bibinfo {author}
  {\bibfnamefont {M.~J.}\ \bibnamefont {Crossley}}, \ and\ \bibinfo {author}
  {\bibfnamefont {T.~W.}\ \bibnamefont {Schmidt}},\ }\href@noop {} {\bibfield
  {journal} {\bibinfo  {journal} {\emph {The Journal of Physical Chemistry
  Letters}}\ }\textbf {\bibinfo {year} {2010}{\natexlab{a}}},\ \emph {\bibinfo
  {volume} {1}},\ \bibinfo {pages} {1795}}\BibitemShut {NoStop}%
\bibitem [{\citenamefont {Tayebjee}\ \emph {et~al.}(2015)\citenamefont
  {Tayebjee}, \citenamefont {McCamey},\ and\ \citenamefont
  {Schmidt}}]{tayebjee2015beyond}%
  \BibitemOpen
  \bibfield  {author} {\bibinfo {author} {\bibfnamefont {M.~J.}\ \bibnamefont
  {Tayebjee}}, \bibinfo {author} {\bibfnamefont {D.~R.}\ \bibnamefont
  {McCamey}}, \ and\ \bibinfo {author} {\bibfnamefont {T.~W.}\ \bibnamefont
  {Schmidt}},\ }\href@noop {} {\bibfield  {journal} {\bibinfo  {journal} {\emph
  {The Journal of Physical Chemistry Letters}}\ }\textbf {\bibinfo {year}
  {2015}},\ \emph {\bibinfo {volume} {6}},\ \bibinfo {pages}
  {2367}}\BibitemShut {NoStop}%
\bibitem [{\citenamefont {Dilbeck}\ \emph {et~al.}(2017)\citenamefont
  {Dilbeck}, \citenamefont {Hill},\ and\ \citenamefont
  {Hanson}}]{dilbeck2017harnessing}%
  \BibitemOpen
  \bibfield  {author} {\bibinfo {author} {\bibfnamefont {T.}~\bibnamefont
  {Dilbeck}}, \bibinfo {author} {\bibfnamefont {S.~P.}\ \bibnamefont {Hill}}, \
  and\ \bibinfo {author} {\bibfnamefont {K.}~\bibnamefont {Hanson}},\
  }\href@noop {} {\bibfield  {journal} {\bibinfo  {journal} {\emph {Journal of
  Materials Chemistry A}}\ }\textbf {\bibinfo {year} {2017}},\ \emph {\bibinfo
  {volume} {5}},\ \bibinfo {pages} {11652}}\BibitemShut {NoStop}%
\bibitem [{\citenamefont {Hill}\ \emph {et~al.}(2016)\citenamefont {Hill},
  \citenamefont {Dilbeck}, \citenamefont {Baduell},\ and\ \citenamefont
  {Hanson}}]{hill2016integrated}%
  \BibitemOpen
  \bibfield  {author} {\bibinfo {author} {\bibfnamefont {S.~P.}\ \bibnamefont
  {Hill}}, \bibinfo {author} {\bibfnamefont {T.}~\bibnamefont {Dilbeck}},
  \bibinfo {author} {\bibfnamefont {E.}~\bibnamefont {Baduell}}, \ and\
  \bibinfo {author} {\bibfnamefont {K.}~\bibnamefont {Hanson}},\ }\href@noop {}
  {\bibfield  {journal} {\bibinfo  {journal} {\emph {ACS Energy Lett.}}\
  }\textbf {\bibinfo {year} {2016}},\ \emph {\bibinfo {volume} {1}},\ \bibinfo
  {pages} {3}}\BibitemShut {NoStop}%
\bibitem [{\citenamefont {Hill}\ \emph {et~al.}(2015)\citenamefont {Hill},
  \citenamefont {Banerjee}, \citenamefont {Dilbeck},\ and\ \citenamefont
  {Hanson}}]{hill2015photon}%
  \BibitemOpen
  \bibfield  {author} {\bibinfo {author} {\bibfnamefont {S.~P.}\ \bibnamefont
  {Hill}}, \bibinfo {author} {\bibfnamefont {T.}~\bibnamefont {Banerjee}},
  \bibinfo {author} {\bibfnamefont {T.}~\bibnamefont {Dilbeck}}, \ and\
  \bibinfo {author} {\bibfnamefont {K.}~\bibnamefont {Hanson}},\ }\href@noop {}
  {\bibfield  {journal} {\bibinfo  {journal} {\emph {J. Phys. Chem. Lett.}}\
  }\textbf {\bibinfo {year} {2015}},\ \emph {\bibinfo {volume} {6}},\ \bibinfo
  {pages} {4510}}\BibitemShut {NoStop}%
\bibitem [{\citenamefont {Gray}\ \emph {et~al.}(2017)\citenamefont {Gray},
  \citenamefont {Dreos}, \citenamefont {Erhart}, \citenamefont {Albinsson},
  \citenamefont {Moth-Poulsen},\ and\ \citenamefont
  {Abrahamsson}}]{gray2017loss}%
  \BibitemOpen
  \bibfield  {author} {\bibinfo {author} {\bibfnamefont {V.}~\bibnamefont
  {Gray}}, \bibinfo {author} {\bibfnamefont {A.}~\bibnamefont {Dreos}},
  \bibinfo {author} {\bibfnamefont {P.}~\bibnamefont {Erhart}}, \bibinfo
  {author} {\bibfnamefont {B.}~\bibnamefont {Albinsson}}, \bibinfo {author}
  {\bibfnamefont {K.}~\bibnamefont {Moth-Poulsen}}, \ and\ \bibinfo {author}
  {\bibfnamefont {M.}~\bibnamefont {Abrahamsson}},\ }\href@noop {} {\bibfield
  {journal} {\bibinfo  {journal} {\emph {Physical Chemistry Chemical Physics}}\
  }\textbf {\bibinfo {year} {2017}},\ \emph {\bibinfo {volume} {19}},\ \bibinfo
  {pages} {10931}}\BibitemShut {NoStop}%
\bibitem [{\citenamefont {Peng}\ \emph {et~al.}(2016)\citenamefont {Peng},
  \citenamefont {Guo}, \citenamefont {Jiang}, \citenamefont {Zhao},\ and\
  \citenamefont {Ma}}]{peng2016developing}%
  \BibitemOpen
  \bibfield  {author} {\bibinfo {author} {\bibfnamefont {J.}~\bibnamefont
  {Peng}}, \bibinfo {author} {\bibfnamefont {X.}~\bibnamefont {Guo}}, \bibinfo
  {author} {\bibfnamefont {X.}~\bibnamefont {Jiang}}, \bibinfo {author}
  {\bibfnamefont {D.}~\bibnamefont {Zhao}}, \ and\ \bibinfo {author}
  {\bibfnamefont {Y.}~\bibnamefont {Ma}},\ }\href@noop {} {\bibfield  {journal}
  {\bibinfo  {journal} {\emph {Chemical Science}}\ }\textbf {\bibinfo {year}
  {2016}},\ \emph {\bibinfo {volume} {7}},\ \bibinfo {pages}
  {1233}}\BibitemShut {NoStop}%
\bibitem [{\citenamefont {Held}(2016)}]{held2016introduction}%
  \BibitemOpen
  \bibfield  {author} {\bibinfo {author} {\bibfnamefont {G.}~\bibnamefont
  {Held}},\ }\href@noop {} {\emph {\bibinfo {title} {Introduction to light
  emitting diode technology and applications}}}\ (\bibinfo  {publisher} {CRC
  press},\ \bibinfo {year} {2016})\BibitemShut {NoStop}%
\bibitem [{\citenamefont {Muramoto}\ \emph {et~al.}(2014)\citenamefont
  {Muramoto}, \citenamefont {Kimura},\ and\ \citenamefont
  {Nouda}}]{muramoto2014development}%
  \BibitemOpen
  \bibfield  {author} {\bibinfo {author} {\bibfnamefont {Y.}~\bibnamefont
  {Muramoto}}, \bibinfo {author} {\bibfnamefont {M.}~\bibnamefont {Kimura}}, \
  and\ \bibinfo {author} {\bibfnamefont {S.}~\bibnamefont {Nouda}},\
  }\href@noop {} {\bibfield  {journal} {\bibinfo  {journal} {\emph
  {Semiconductor Science and Technology}}\ }\textbf {\bibinfo {year} {2014}},\
  \emph {\bibinfo {volume} {29}},\ \bibinfo {pages} {084004}}\BibitemShut
  {NoStop}%
\bibitem [{\citenamefont {Frazer}\ \emph {et~al.}(2017)\citenamefont {Frazer},
  \citenamefont {Gallaher},\ and\ \citenamefont
  {Schmidt}}]{frazer2017optimizing}%
  \BibitemOpen
  \bibfield  {author} {\bibinfo {author} {\bibfnamefont {L.}~\bibnamefont
  {Frazer}}, \bibinfo {author} {\bibfnamefont {J.~K.}\ \bibnamefont
  {Gallaher}}, \ and\ \bibinfo {author} {\bibfnamefont {T.}~\bibnamefont
  {Schmidt}},\ }\href@noop {} {\bibfield  {journal} {\bibinfo  {journal} {\emph
  {ACS Energy Letters}}\ }\textbf {\bibinfo {year} {2017}},\ \emph {\bibinfo
  {volume} {2}},\ \bibinfo {pages} {1346}}\BibitemShut {NoStop}%
\bibitem [{\citenamefont {Zeng}\ \emph {et~al.}(2017)\citenamefont {Zeng},
  \citenamefont {Chen}, \citenamefont {Yu}, \citenamefont {Yang},\ and\
  \citenamefont {Li}}]{zeng2017molecular}%
  \BibitemOpen
  \bibfield  {author} {\bibinfo {author} {\bibfnamefont {Y.}~\bibnamefont
  {Zeng}}, \bibinfo {author} {\bibfnamefont {J.}~\bibnamefont {Chen}}, \bibinfo
  {author} {\bibfnamefont {T.}~\bibnamefont {Yu}}, \bibinfo {author}
  {\bibfnamefont {G.}~\bibnamefont {Yang}}, \ and\ \bibinfo {author}
  {\bibfnamefont {Y.}~\bibnamefont {Li}},\ }\href@noop {} {\bibfield  {journal}
  {\bibinfo  {journal} {\emph {ACS Energy Letters}}\ }\textbf {\bibinfo {year}
  {2017}},\ \emph {\bibinfo {volume} {2}},\ \bibinfo {pages} {357}}\BibitemShut
  {NoStop}%
\bibitem [{\citenamefont {Pedrini}\ and\ \citenamefont
  {Monguzzi}(2017)}]{pedrini2017recent}%
  \BibitemOpen
  \bibfield  {author} {\bibinfo {author} {\bibfnamefont {J.}~\bibnamefont
  {Pedrini}}\ and\ \bibinfo {author} {\bibfnamefont {A.}~\bibnamefont
  {Monguzzi}},\ }\href@noop {} {\bibfield  {journal} {\bibinfo  {journal}
  {\emph {Journal of Photonics for Energy}}\ }\textbf {\bibinfo {year}
  {2017}},\ \emph {\bibinfo {volume} {8}},\ \bibinfo {pages}
  {022005}}\BibitemShut {NoStop}%
\bibitem [{\citenamefont {von Reventlow}\ \emph {et~al.}(2018)\citenamefont
  {von Reventlow}, \citenamefont {Bremer}, \citenamefont {Ebenhoch},
  \citenamefont {Gerken}, \citenamefont {Schmidt},\ and\ \citenamefont
  {Colsmann}}]{von2018add}%
  \BibitemOpen
  \bibfield  {author} {\bibinfo {author} {\bibfnamefont {L.~G.}\ \bibnamefont
  {von Reventlow}}, \bibinfo {author} {\bibfnamefont {M.}~\bibnamefont
  {Bremer}}, \bibinfo {author} {\bibfnamefont {B.}~\bibnamefont {Ebenhoch}},
  \bibinfo {author} {\bibfnamefont {M.}~\bibnamefont {Gerken}}, \bibinfo
  {author} {\bibfnamefont {T.~W.}\ \bibnamefont {Schmidt}}, \ and\ \bibinfo
  {author} {\bibfnamefont {A.}~\bibnamefont {Colsmann}},\ }\href@noop {}
  {\bibfield  {journal} {\bibinfo  {journal} {\emph {Journal of Materials
  Chemistry C}}\ }\textbf {\bibinfo {year} {2018}},\ \emph {\bibinfo {volume}
  {6}},\ \bibinfo {pages} {3845}}\BibitemShut {NoStop}%
\bibitem [{\citenamefont {Wanvik}(2009)}]{wanvik2009effects}%
  \BibitemOpen
  \bibfield  {author} {\bibinfo {author} {\bibfnamefont {P.~O.}\ \bibnamefont
  {Wanvik}},\ }\href@noop {} {\bibfield  {journal} {\bibinfo  {journal} {\emph
  {Accident Analysis \& Prevention}}\ }\textbf {\bibinfo {year} {2009}},\ \emph
  {\bibinfo {volume} {41}},\ \bibinfo {pages} {123}}\BibitemShut {NoStop}%
\bibitem [{\citenamefont {Gaston}\ \emph {et~al.}(2014)\citenamefont {Gaston},
  \citenamefont {Gaston}, \citenamefont {Bennie},\ and\ \citenamefont
  {Hopkins}}]{gaston2014benefits}%
  \BibitemOpen
  \bibfield  {author} {\bibinfo {author} {\bibfnamefont {K.~J.}\ \bibnamefont
  {Gaston}}, \bibinfo {author} {\bibfnamefont {S.}~\bibnamefont {Gaston}},
  \bibinfo {author} {\bibfnamefont {J.}~\bibnamefont {Bennie}}, \ and\ \bibinfo
  {author} {\bibfnamefont {J.}~\bibnamefont {Hopkins}},\ }\href@noop {}
  {\bibfield  {journal} {\bibinfo  {journal} {\emph {Environmental Reviews}}\
  }\textbf {\bibinfo {year} {2014}},\ \emph {\bibinfo {volume} {23}},\ \bibinfo
  {pages} {14}}\BibitemShut {NoStop}%
\bibitem [{eia(2017)}]{eia2017}%
  \BibitemOpen
  \href {https://www.eia.gov/tools/faqs/faq.php?id=99&t=3} {\emph {\bibinfo
  {title} {How much electricity is used for lighting in the United States?}}}\
  (\bibinfo  {publisher} {U.S. Energy Information Administration},\ \bibinfo
  {year} {2017})\BibitemShut {NoStop}%
\bibitem [{\citenamefont {Zhou}\ \emph {et~al.}(2018)\citenamefont {Zhou},
  \citenamefont {Ayad}, \citenamefont {Ruchlin}, \citenamefont {Posey},
  \citenamefont {Hill}, \citenamefont {Wu},\ and\ \citenamefont
  {Hanson}}]{zhou2018examining}%
  \BibitemOpen
  \bibfield  {author} {\bibinfo {author} {\bibfnamefont {Y.}~\bibnamefont
  {Zhou}}, \bibinfo {author} {\bibfnamefont {S.}~\bibnamefont {Ayad}}, \bibinfo
  {author} {\bibfnamefont {C.}~\bibnamefont {Ruchlin}}, \bibinfo {author}
  {\bibfnamefont {V.}~\bibnamefont {Posey}}, \bibinfo {author} {\bibfnamefont
  {S.~P.}\ \bibnamefont {Hill}}, \bibinfo {author} {\bibfnamefont
  {Q.}~\bibnamefont {Wu}}, \ and\ \bibinfo {author} {\bibfnamefont
  {K.}~\bibnamefont {Hanson}},\ }\href@noop {} {\bibfield  {journal} {\bibinfo
  {journal} {\emph {Physical Chemistry Chemical Physics}}\ }\textbf {\bibinfo
  {year} {2018}},\ \emph {\bibinfo {volume} {20}}}\BibitemShut {NoStop}%
\bibitem [{\citenamefont {Pust}\ \emph {et~al.}(2015)\citenamefont {Pust},
  \citenamefont {Schmidt},\ and\ \citenamefont {Schnick}}]{pust2015revolution}%
  \BibitemOpen
  \bibfield  {author} {\bibinfo {author} {\bibfnamefont {P.}~\bibnamefont
  {Pust}}, \bibinfo {author} {\bibfnamefont {P.~J.}\ \bibnamefont {Schmidt}}, \
  and\ \bibinfo {author} {\bibfnamefont {W.}~\bibnamefont {Schnick}},\
  }\href@noop {} {\bibfield  {journal} {\bibinfo  {journal} {\emph {Nature
  Materials}}\ }\textbf {\bibinfo {year} {2015}},\ \emph {\bibinfo {volume}
  {14}},\ \bibinfo {pages} {454}}\BibitemShut {NoStop}%
\bibitem [{\citenamefont {Koyama}(2006)}]{koyama2006recent}%
  \BibitemOpen
  \bibfield  {author} {\bibinfo {author} {\bibfnamefont {F.}~\bibnamefont
  {Koyama}},\ }\href@noop {} {\bibfield  {journal} {\bibinfo  {journal} {\emph
  {Journal of Lightwave Technology}}\ }\textbf {\bibinfo {year} {2006}},\ \emph
  {\bibinfo {volume} {24}},\ \bibinfo {pages} {4502}}\BibitemShut {NoStop}%
\bibitem [{\citenamefont {Guo}\ \emph {et~al.}(2012)\citenamefont {Guo},
  \citenamefont {Wu}, \citenamefont {Guo},\ and\ \citenamefont
  {Zhao}}]{guo2012room}%
  \BibitemOpen
  \bibfield  {author} {\bibinfo {author} {\bibfnamefont {S.}~\bibnamefont
  {Guo}}, \bibinfo {author} {\bibfnamefont {W.}~\bibnamefont {Wu}}, \bibinfo
  {author} {\bibfnamefont {H.}~\bibnamefont {Guo}}, \ and\ \bibinfo {author}
  {\bibfnamefont {J.}~\bibnamefont {Zhao}},\ }\href@noop {} {\bibfield
  {journal} {\bibinfo  {journal} {\emph {The Journal of Organic Chemistry}}\
  }\textbf {\bibinfo {year} {2012}},\ \emph {\bibinfo {volume} {77}},\ \bibinfo
  {pages} {3933}}\BibitemShut {NoStop}%
\bibitem [{\citenamefont {Cui}\ \emph {et~al.}(2013)\citenamefont {Cui},
  \citenamefont {Zhao}, \citenamefont {Yang},\ and\ \citenamefont
  {Sun}}]{cui2013zinc}%
  \BibitemOpen
  \bibfield  {author} {\bibinfo {author} {\bibfnamefont {X.}~\bibnamefont
  {Cui}}, \bibinfo {author} {\bibfnamefont {J.}~\bibnamefont {Zhao}}, \bibinfo
  {author} {\bibfnamefont {P.}~\bibnamefont {Yang}}, \ and\ \bibinfo {author}
  {\bibfnamefont {J.}~\bibnamefont {Sun}},\ }\href@noop {} {\bibfield
  {journal} {\bibinfo  {journal} {\emph {Chemical Communications}}\ }\textbf
  {\bibinfo {year} {2013}},\ \emph {\bibinfo {volume} {49}},\ \bibinfo {pages}
  {10221}}\BibitemShut {NoStop}%
\bibitem [{\citenamefont {Islangulov}\ \emph {et~al.}(2005)\citenamefont
  {Islangulov}, \citenamefont {Kozlov},\ and\ \citenamefont
  {Castellano}}]{islangulov2005low}%
  \BibitemOpen
  \bibfield  {author} {\bibinfo {author} {\bibfnamefont {R.~R.}\ \bibnamefont
  {Islangulov}}, \bibinfo {author} {\bibfnamefont {D.~V.}\ \bibnamefont
  {Kozlov}}, \ and\ \bibinfo {author} {\bibfnamefont {F.~N.}\ \bibnamefont
  {Castellano}},\ }\href@noop {} {\bibfield  {journal} {\bibinfo  {journal}
  {\emph {Chemical Communications}}\ }\textbf {\bibinfo {year} {2005}},\ \emph
  {\bibinfo {volume} {2005}},\ \bibinfo {pages} {3776}}\BibitemShut {NoStop}%
\bibitem [{\citenamefont {Wu}\ \emph {et~al.}(2011)\citenamefont {Wu},
  \citenamefont {Guo}, \citenamefont {Wu}, \citenamefont {Ji},\ and\
  \citenamefont {Zhao}}]{wu2011organic}%
  \BibitemOpen
  \bibfield  {author} {\bibinfo {author} {\bibfnamefont {W.}~\bibnamefont
  {Wu}}, \bibinfo {author} {\bibfnamefont {H.}~\bibnamefont {Guo}}, \bibinfo
  {author} {\bibfnamefont {W.}~\bibnamefont {Wu}}, \bibinfo {author}
  {\bibfnamefont {S.}~\bibnamefont {Ji}}, \ and\ \bibinfo {author}
  {\bibfnamefont {J.}~\bibnamefont {Zhao}},\ }\href@noop {} {\bibfield
  {journal} {\bibinfo  {journal} {\emph {Journal of Organic Chemistry}}\
  }\textbf {\bibinfo {year} {2011}},\ \emph {\bibinfo {volume} {76}},\ \bibinfo
  {pages} {7056}}\BibitemShut {NoStop}%
\bibitem [{\citenamefont {Yu}\ \emph {et~al.}(2015)\citenamefont {Yu},
  \citenamefont {Cao}, \citenamefont {Chen}, \citenamefont {Ayres},\ and\
  \citenamefont {Zhang}}]{yu2015triplet}%
  \BibitemOpen
  \bibfield  {author} {\bibinfo {author} {\bibfnamefont {X.}~\bibnamefont
  {Yu}}, \bibinfo {author} {\bibfnamefont {X.}~\bibnamefont {Cao}}, \bibinfo
  {author} {\bibfnamefont {X.}~\bibnamefont {Chen}}, \bibinfo {author}
  {\bibfnamefont {N.}~\bibnamefont {Ayres}}, \ and\ \bibinfo {author}
  {\bibfnamefont {P.}~\bibnamefont {Zhang}},\ }\href@noop {} {\bibfield
  {journal} {\bibinfo  {journal} {\emph {Chemical Communications}}\ }\textbf
  {\bibinfo {year} {2015}},\ \emph {\bibinfo {volume} {51}},\ \bibinfo {pages}
  {588}}\BibitemShut {NoStop}%
\bibitem [{\citenamefont {Penconi}\ \emph {et~al.}(2013)\citenamefont
  {Penconi}, \citenamefont {Ortica}, \citenamefont {Elisei},\ and\
  \citenamefont {Gentili}}]{penconi2013new}%
  \BibitemOpen
  \bibfield  {author} {\bibinfo {author} {\bibfnamefont {M.}~\bibnamefont
  {Penconi}}, \bibinfo {author} {\bibfnamefont {F.}~\bibnamefont {Ortica}},
  \bibinfo {author} {\bibfnamefont {F.}~\bibnamefont {Elisei}}, \ and\ \bibinfo
  {author} {\bibfnamefont {P.~L.}\ \bibnamefont {Gentili}},\ }\href@noop {}
  {\bibfield  {journal} {\bibinfo  {journal} {\emph {Journal of Luminescence}}\
  }\textbf {\bibinfo {year} {2013}},\ \emph {\bibinfo {volume} {135}},\
  \bibinfo {pages} {265}}\BibitemShut {NoStop}%
\bibitem [{\citenamefont {Deng}\ \emph {et~al.}(2015)\citenamefont {Deng},
  \citenamefont {Francis}, \citenamefont {Weare},\ and\ \citenamefont
  {Castellano}}]{deng2015photochemical}%
  \BibitemOpen
  \bibfield  {author} {\bibinfo {author} {\bibfnamefont {F.}~\bibnamefont
  {Deng}}, \bibinfo {author} {\bibfnamefont {A.}~\bibnamefont {Francis}},
  \bibinfo {author} {\bibfnamefont {W.}~\bibnamefont {Weare}}, \ and\ \bibinfo
  {author} {\bibfnamefont {F.}~\bibnamefont {Castellano}},\ }\href@noop {}
  {\bibfield  {journal} {\bibinfo  {journal} {\emph {Photochemical \&
  Photobiological Sciences}}\ }\textbf {\bibinfo {year} {2015}},\ \emph
  {\bibinfo {volume} {14}},\ \bibinfo {pages} {1265}}\BibitemShut {NoStop}%
\bibitem [{\citenamefont {Wu}\ \emph {et~al.}(2015)\citenamefont {Wu},
  \citenamefont {Congreve},\ and\ \citenamefont {Baldo}}]{wu2015solid}%
  \BibitemOpen
  \bibfield  {author} {\bibinfo {author} {\bibfnamefont {T.~C.}\ \bibnamefont
  {Wu}}, \bibinfo {author} {\bibfnamefont {D.~N.}\ \bibnamefont {Congreve}}, \
  and\ \bibinfo {author} {\bibfnamefont {M.~A.}\ \bibnamefont {Baldo}},\
  }\href@noop {} {\bibfield  {journal} {\bibinfo  {journal} {\emph {Applied
  Physics Lett.}}\ }\textbf {\bibinfo {year} {2015}},\ \emph {\bibinfo {volume}
  {107}},\ \bibinfo {pages} {031103}}\BibitemShut {NoStop}%
\bibitem [{\citenamefont {Baluschev}\ \emph {et~al.}(2007)\citenamefont
  {Baluschev}, \citenamefont {Yakutkin}, \citenamefont {Wegner}, \citenamefont
  {Miteva}, \citenamefont {Nelles}, \citenamefont {Yasuda}, \citenamefont
  {Chernov}, \citenamefont {Aleshchenkov},\ and\ \citenamefont
  {Cheprakov}}]{Baluschev_2007}%
  \BibitemOpen
  \bibfield  {author} {\bibinfo {author} {\bibfnamefont {S.}~\bibnamefont
  {Baluschev}}, \bibinfo {author} {\bibfnamefont {V.}~\bibnamefont {Yakutkin}},
  \bibinfo {author} {\bibfnamefont {G.}~\bibnamefont {Wegner}}, \bibinfo
  {author} {\bibfnamefont {T.}~\bibnamefont {Miteva}}, \bibinfo {author}
  {\bibfnamefont {G.}~\bibnamefont {Nelles}}, \bibinfo {author} {\bibfnamefont
  {A.}~\bibnamefont {Yasuda}}, \bibinfo {author} {\bibfnamefont
  {S.}~\bibnamefont {Chernov}}, \bibinfo {author} {\bibfnamefont
  {S.}~\bibnamefont {Aleshchenkov}}, \ and\ \bibinfo {author} {\bibfnamefont
  {A.}~\bibnamefont {Cheprakov}},\ }\href@noop {} {\bibfield  {journal}
  {\bibinfo  {journal} {\emph {Applied Phys. Lett.}}\ }\textbf {\bibinfo {year}
  {2007}},\ \emph {\bibinfo {volume} {90}},\ \bibinfo {eid}
  {181103}}\BibitemShut {NoStop}%
\bibitem [{\citenamefont {Cheng}\ \emph
  {et~al.}(2010{\natexlab{b}})\citenamefont {Cheng}, \citenamefont {Khoury},
  \citenamefont {Clady}, \citenamefont {Tayebjee}, \citenamefont
  {Ekins-Daukes}, \citenamefont {Crossley},\ and\ \citenamefont
  {Schmidt}}]{cheng2010efficiency}%
  \BibitemOpen
  \bibfield  {author} {\bibinfo {author} {\bibfnamefont {Y.~Y.}\ \bibnamefont
  {Cheng}}, \bibinfo {author} {\bibfnamefont {T.}~\bibnamefont {Khoury}},
  \bibinfo {author} {\bibfnamefont {R.~G.}\ \bibnamefont {Clady}}, \bibinfo
  {author} {\bibfnamefont {M.~J.}\ \bibnamefont {Tayebjee}}, \bibinfo {author}
  {\bibfnamefont {N.}~\bibnamefont {Ekins-Daukes}}, \bibinfo {author}
  {\bibfnamefont {M.~J.}\ \bibnamefont {Crossley}}, \ and\ \bibinfo {author}
  {\bibfnamefont {T.~W.}\ \bibnamefont {Schmidt}},\ }\href@noop {} {\bibfield
  {journal} {\bibinfo  {journal} {\emph {Physical Chemistry Chemical Physics}}\
  }\textbf {\bibinfo {year} {2010}{\natexlab{b}}},\ \emph {\bibinfo {volume}
  {12}},\ \bibinfo {pages} {66}}\BibitemShut {NoStop}%
\bibitem [{\citenamefont {O'Hara}\ \emph {et~al.}(1999)\citenamefont {O'Hara},
  \citenamefont {Gullingsrud},\ and\ \citenamefont {Wolfe}}]{o1999auger}%
  \BibitemOpen
  \bibfield  {author} {\bibinfo {author} {\bibfnamefont {K.}~\bibnamefont
  {O'Hara}}, \bibinfo {author} {\bibfnamefont {J.}~\bibnamefont {Gullingsrud}},
  \ and\ \bibinfo {author} {\bibfnamefont {J.}~\bibnamefont {Wolfe}},\
  }\href@noop {} {\bibfield  {journal} {\bibinfo  {journal} {\emph {Physical
  Review B}}\ }\textbf {\bibinfo {year} {1999}},\ \emph {\bibinfo {volume}
  {60}},\ \bibinfo {pages} {10872}}\BibitemShut {NoStop}%
\bibitem [{\citenamefont {Frazer}\ \emph {et~al.}(2013)\citenamefont {Frazer},
  \citenamefont {Schaller},\ and\ \citenamefont
  {Ketterson}}]{laszlo2013unexpectedly}%
  \BibitemOpen
  \bibfield  {author} {\bibinfo {author} {\bibfnamefont {L.}~\bibnamefont
  {Frazer}}, \bibinfo {author} {\bibfnamefont {R.~D.}\ \bibnamefont
  {Schaller}}, \ and\ \bibinfo {author} {\bibfnamefont {J.}~\bibnamefont
  {Ketterson}},\ }\href@noop {} {\bibfield  {journal} {\bibinfo  {journal}
  {\emph {Solid State Communications}}\ }\textbf {\bibinfo {year} {2013}},\
  \emph {\bibinfo {volume} {170}},\ \bibinfo {pages} {34}}\BibitemShut
  {NoStop}%
\bibitem [{\citenamefont {Cheng}\ \emph {et~al.}(2011)\citenamefont {Cheng},
  \citenamefont {F\"uckel}, \citenamefont {Khoury}, \citenamefont {Clady},
  \citenamefont {Ekins-Daukes}, \citenamefont {Crossley},\ and\ \citenamefont
  {Schmidt}}]{cheng2011entropically}%
  \BibitemOpen
  \bibfield  {author} {\bibinfo {author} {\bibfnamefont {Y.~Y.}\ \bibnamefont
  {Cheng}}, \bibinfo {author} {\bibfnamefont {B.}~\bibnamefont {F\"uckel}},
  \bibinfo {author} {\bibfnamefont {T.}~\bibnamefont {Khoury}}, \bibinfo
  {author} {\bibfnamefont {R.~G.}\ \bibnamefont {Clady}}, \bibinfo {author}
  {\bibfnamefont {N.}~\bibnamefont {Ekins-Daukes}}, \bibinfo {author}
  {\bibfnamefont {M.~J.}\ \bibnamefont {Crossley}}, \ and\ \bibinfo {author}
  {\bibfnamefont {T.~W.}\ \bibnamefont {Schmidt}},\ }\href@noop {} {\bibfield
  {journal} {\bibinfo  {journal} {\emph {The Journal of Physical Chemistry A}}\
  }\textbf {\bibinfo {year} {2011}},\ \emph {\bibinfo {volume} {115}},\
  \bibinfo {pages} {1047}}\BibitemShut {NoStop}%
\bibitem [{\citenamefont {Dover}\ \emph {et~al.}(2018)\citenamefont {Dover},
  \citenamefont {Gallaher}, \citenamefont {Frazer}, \citenamefont {Tapping},
  \citenamefont {Petty~II}, \citenamefont {Crossley}, \citenamefont {Anthony},
  \citenamefont {Kee},\ and\ \citenamefont {Schmidt}}]{dover2018endothermic}%
  \BibitemOpen
  \bibfield  {author} {\bibinfo {author} {\bibfnamefont {C.~B.}\ \bibnamefont
  {Dover}}, \bibinfo {author} {\bibfnamefont {J.~K.}\ \bibnamefont {Gallaher}},
  \bibinfo {author} {\bibfnamefont {L.}~\bibnamefont {Frazer}}, \bibinfo
  {author} {\bibfnamefont {P.~C.}\ \bibnamefont {Tapping}}, \bibinfo {author}
  {\bibfnamefont {A.~J.}\ \bibnamefont {Petty~II}}, \bibinfo {author}
  {\bibfnamefont {M.~J.}\ \bibnamefont {Crossley}}, \bibinfo {author}
  {\bibfnamefont {J.~E.}\ \bibnamefont {Anthony}}, \bibinfo {author}
  {\bibfnamefont {T.~W.}\ \bibnamefont {Kee}}, \ and\ \bibinfo {author}
  {\bibfnamefont {T.~W.}\ \bibnamefont {Schmidt}},\ }\href@noop {} {\bibfield
  {journal} {\bibinfo  {journal} {\emph {Nature Chemistry}}\ }\textbf {\bibinfo
  {year} {2018}},\ \emph {\bibinfo {volume} {10}},\ \bibinfo {pages}
  {305}}\BibitemShut {NoStop}%
\bibitem [{\citenamefont {Ware}\ and\ \citenamefont
  {Baldwin}(1965)}]{ware1965effect}%
  \BibitemOpen
  \bibfield  {author} {\bibinfo {author} {\bibfnamefont {W.~R.}\ \bibnamefont
  {Ware}}\ and\ \bibinfo {author} {\bibfnamefont {B.~A.}\ \bibnamefont
  {Baldwin}},\ }\href@noop {} {\bibfield  {journal} {\bibinfo  {journal} {\emph
  {The Journal of Chemical Physics}}\ }\textbf {\bibinfo {year} {1965}},\ \emph
  {\bibinfo {volume} {43}},\ \bibinfo {pages} {1194}}\BibitemShut {NoStop}%
\bibitem [{\citenamefont {Bennett}\ and\ \citenamefont
  {McCartin}(1966)}]{bennett1966radiationless}%
  \BibitemOpen
  \bibfield  {author} {\bibinfo {author} {\bibfnamefont {R.}~\bibnamefont
  {Bennett}}\ and\ \bibinfo {author} {\bibfnamefont {P.~J.}\ \bibnamefont
  {McCartin}},\ }\href@noop {} {\bibfield  {journal} {\bibinfo  {journal}
  {\emph {The Journal of Chemical Physics}}\ }\textbf {\bibinfo {year}
  {1966}},\ \emph {\bibinfo {volume} {44}},\ \bibinfo {pages}
  {1969}}\BibitemShut {NoStop}%
\bibitem [{\citenamefont {Lim}\ \emph {et~al.}(1966)\citenamefont {Lim},
  \citenamefont {Laposa},\ and\ \citenamefont {Jack}}]{lim1966temperature}%
  \BibitemOpen
  \bibfield  {author} {\bibinfo {author} {\bibfnamefont {E.}~\bibnamefont
  {Lim}}, \bibinfo {author} {\bibfnamefont {J.~D.}\ \bibnamefont {Laposa}}, \
  and\ \bibinfo {author} {\bibfnamefont {M.}~\bibnamefont {Jack}},\ }\href@noop
  {} {\bibfield  {journal} {\bibinfo  {journal} {\emph {Journal of Molecular
  Spectroscopy}}\ }\textbf {\bibinfo {year} {1966}},\ \emph {\bibinfo {volume}
  {19}},\ \bibinfo {pages} {412}}\BibitemShut {NoStop}%
\bibitem [{\citenamefont {Gillispie}\ and\ \citenamefont
  {Lim}(1976)}]{gillispie1976t}%
  \BibitemOpen
  \bibfield  {author} {\bibinfo {author} {\bibfnamefont {G.~D.}\ \bibnamefont
  {Gillispie}}\ and\ \bibinfo {author} {\bibfnamefont {E.}~\bibnamefont
  {Lim}},\ }\href@noop {} {\bibfield  {journal} {\bibinfo  {journal} {\emph
  {The Journal of Chemical Physics}}\ }\textbf {\bibinfo {year} {1976}},\ \emph
  {\bibinfo {volume} {65}},\ \bibinfo {pages} {2022}}\BibitemShut {NoStop}%
\bibitem [{\citenamefont {Wang}\ \emph {et~al.}(2004)\citenamefont {Wang},
  \citenamefont {Del~Guerzo},\ and\ \citenamefont
  {Schmehl}}]{wang2004photophysical}%
  \BibitemOpen
  \bibfield  {author} {\bibinfo {author} {\bibfnamefont {X.-Y.}\ \bibnamefont
  {Wang}}, \bibinfo {author} {\bibfnamefont {A.}~\bibnamefont {Del~Guerzo}}, \
  and\ \bibinfo {author} {\bibfnamefont {R.~H.}\ \bibnamefont {Schmehl}},\
  }\href@noop {} {\bibfield  {journal} {\bibinfo  {journal} {\emph {Journal of
  Photochemistry and Photobiology C: Photochemistry Reviews}}\ }\textbf
  {\bibinfo {year} {2004}},\ \emph {\bibinfo {volume} {5}},\ \bibinfo {pages}
  {55}}\BibitemShut {NoStop}%
\bibitem [{\citenamefont {Perun}\ \emph {et~al.}(2008)\citenamefont {Perun},
  \citenamefont {Tatchen},\ and\ \citenamefont {Marian}}]{perun2008singlet}%
  \BibitemOpen
  \bibfield  {author} {\bibinfo {author} {\bibfnamefont {S.}~\bibnamefont
  {Perun}}, \bibinfo {author} {\bibfnamefont {J.}~\bibnamefont {Tatchen}}, \
  and\ \bibinfo {author} {\bibfnamefont {C.~M.}\ \bibnamefont {Marian}},\
  }\href@noop {} {\bibfield  {journal} {\bibinfo  {journal} {\emph
  {ChemPhysChem}}\ }\textbf {\bibinfo {year} {2008}},\ \emph {\bibinfo {volume}
  {9}},\ \bibinfo {pages} {282}}\BibitemShut {NoStop}%
\bibitem [{\citenamefont {Hoseinkhani}\ \emph {et~al.}(2015)\citenamefont
  {Hoseinkhani}, \citenamefont {Tubino}, \citenamefont {Meinardi},\ and\
  \citenamefont {Monguzzi}}]{hoseinkhani2015achieving}%
  \BibitemOpen
  \bibfield  {author} {\bibinfo {author} {\bibfnamefont {S.}~\bibnamefont
  {Hoseinkhani}}, \bibinfo {author} {\bibfnamefont {R.}~\bibnamefont {Tubino}},
  \bibinfo {author} {\bibfnamefont {F.}~\bibnamefont {Meinardi}}, \ and\
  \bibinfo {author} {\bibfnamefont {A.}~\bibnamefont {Monguzzi}},\ }\href@noop
  {} {\bibfield  {journal} {\bibinfo  {journal} {\emph {Physical Chemistry
  Chemical Physics}}\ }\textbf {\bibinfo {year} {2015}},\ \emph {\bibinfo
  {volume} {17}},\ \bibinfo {pages} {4020}}\BibitemShut {NoStop}%
\bibitem [{\citenamefont {Schmidt}\ and\ \citenamefont
  {Castellano}(2014)}]{schmidt2014photochemical}%
  \BibitemOpen
  \bibfield  {author} {\bibinfo {author} {\bibfnamefont {T.~W.}\ \bibnamefont
  {Schmidt}}\ and\ \bibinfo {author} {\bibfnamefont {F.~N.}\ \bibnamefont
  {Castellano}},\ }\href@noop {} {\bibfield  {journal} {\bibinfo  {journal}
  {\emph {The Journal of Physical Chemistry Letters}}\ }\textbf {\bibinfo
  {year} {2014}},\ \emph {\bibinfo {volume} {5}},\ \bibinfo {pages}
  {4062}}\BibitemShut {NoStop}%
\bibitem [{\citenamefont {Gholizadeh}\ \emph {et~al.}(2018)\citenamefont
  {Gholizadeh}, \citenamefont {Frazer}, \citenamefont {MacQueen}, \citenamefont
  {Gallaher},\ and\ \citenamefont {Schmidt}}]{gholizadeh2018photochemical}%
  \BibitemOpen
  \bibfield  {author} {\bibinfo {author} {\bibfnamefont {E.~M.}\ \bibnamefont
  {Gholizadeh}}, \bibinfo {author} {\bibfnamefont {L.}~\bibnamefont {Frazer}},
  \bibinfo {author} {\bibfnamefont {R.}~\bibnamefont {MacQueen}}, \bibinfo
  {author} {\bibfnamefont {J.}~\bibnamefont {Gallaher}}, \ and\ \bibinfo
  {author} {\bibfnamefont {T.~W.}\ \bibnamefont {Schmidt}},\ }\href@noop {}
  {\bibfield  {journal} {\bibinfo  {journal} {\emph {Physical Chemistry
  Chemical Physics}}\ }\textbf {\bibinfo {year} {2018}},\ \emph {\bibinfo
  {volume} {20}},\ \bibinfo {pages} {19500}}\BibitemShut {NoStop}%
\bibitem [{\citenamefont {Siebrand}(1967)}]{siebrand1967radiationless}%
  \BibitemOpen
  \bibfield  {author} {\bibinfo {author} {\bibfnamefont {W.}~\bibnamefont
  {Siebrand}},\ }\href@noop {} {\bibfield  {journal} {\bibinfo  {journal}
  {\emph {The Journal of Chemical Physics}}\ }\textbf {\bibinfo {year}
  {1967}},\ \emph {\bibinfo {volume} {47}},\ \bibinfo {pages}
  {2411}}\BibitemShut {NoStop}%
\bibitem [{\citenamefont {Hui}\ and\ \citenamefont
  {Qin}(2009)}]{hui2009general}%
  \BibitemOpen
  \bibfield  {author} {\bibinfo {author} {\bibfnamefont {S.}~\bibnamefont
  {Hui}}\ and\ \bibinfo {author} {\bibfnamefont {Y.}~\bibnamefont {Qin}},\
  }\href@noop {} {\bibfield  {journal} {\bibinfo  {journal} {\emph {IEEE
  Transactions on Power Electronics}}\ }\textbf {\bibinfo {year} {2009}},\
  \emph {\bibinfo {volume} {24}},\ \bibinfo {pages} {1967}}\BibitemShut
  {NoStop}%
\bibitem [{\citenamefont {Qin}\ \emph {et~al.}(2009)\citenamefont {Qin},
  \citenamefont {Lin},\ and\ \citenamefont {Hui}}]{qin2009simple}%
  \BibitemOpen
  \bibfield  {author} {\bibinfo {author} {\bibfnamefont {Y.}~\bibnamefont
  {Qin}}, \bibinfo {author} {\bibfnamefont {D.}~\bibnamefont {Lin}}, \ and\
  \bibinfo {author} {\bibfnamefont {S.}~\bibnamefont {Hui}},\ }in\ \href@noop
  {} {\emph {\bibinfo {booktitle} {Applied Power Electronics Conference and
  Exposition, 2009. APEC 2009. Twenty-Fourth Annual IEEE}}}\ (\bibinfo
  {organization} {IEEE},\ \bibinfo {year} {2009})\ pp.\ \bibinfo {pages}
  {152--158}\BibitemShut {NoStop}%
\bibitem [{\citenamefont {MacQueen}\ \emph {et~al.}(2014)\citenamefont
  {MacQueen}, \citenamefont {Cheng}, \citenamefont {Danos}, \citenamefont
  {Lips},\ and\ \citenamefont {Schmidt}}]{macqueen2014action}%
  \BibitemOpen
  \bibfield  {author} {\bibinfo {author} {\bibfnamefont {R.~W.}\ \bibnamefont
  {MacQueen}}, \bibinfo {author} {\bibfnamefont {Y.~Y.}\ \bibnamefont {Cheng}},
  \bibinfo {author} {\bibfnamefont {A.~N.}\ \bibnamefont {Danos}}, \bibinfo
  {author} {\bibfnamefont {K.}~\bibnamefont {Lips}}, \ and\ \bibinfo {author}
  {\bibfnamefont {T.~W.}\ \bibnamefont {Schmidt}},\ }\href@noop {} {\bibfield
  {journal} {\bibinfo  {journal} {\emph {RSC Advances}}\ }\textbf {\bibinfo
  {year} {2014}},\ \emph {\bibinfo {volume} {4}},\ \bibinfo {pages}
  {52749}}\BibitemShut {NoStop}%
\bibitem [{\citenamefont {Dick}\ and\ \citenamefont
  {Nickel}(1983)}]{dick1983accessibility}%
  \BibitemOpen
  \bibfield  {author} {\bibinfo {author} {\bibfnamefont {B.}~\bibnamefont
  {Dick}}\ and\ \bibinfo {author} {\bibfnamefont {B.}~\bibnamefont {Nickel}},\
  }\href@noop {} {\bibfield  {journal} {\bibinfo  {journal} {\emph {Chemical
  Physics}}\ }\textbf {\bibinfo {year} {1983}},\ \emph {\bibinfo {volume}
  {78}},\ \bibinfo {pages} {1}}\BibitemShut {NoStop}%
\bibitem [{\citenamefont {Gray}\ \emph {et~al.}(2014)\citenamefont {Gray},
  \citenamefont {Dzebo}, \citenamefont {Abrahamsson}, \citenamefont
  {Albinsson},\ and\ \citenamefont {Moth-Poulsen}}]{gray2014triplet}%
  \BibitemOpen
  \bibfield  {author} {\bibinfo {author} {\bibfnamefont {V.}~\bibnamefont
  {Gray}}, \bibinfo {author} {\bibfnamefont {D.}~\bibnamefont {Dzebo}},
  \bibinfo {author} {\bibfnamefont {M.}~\bibnamefont {Abrahamsson}}, \bibinfo
  {author} {\bibfnamefont {B.}~\bibnamefont {Albinsson}}, \ and\ \bibinfo
  {author} {\bibfnamefont {K.}~\bibnamefont {Moth-Poulsen}},\ }\href@noop {}
  {\bibfield  {journal} {\bibinfo  {journal} {\emph {Physical Chemistry
  Chemical Physics}}\ }\textbf {\bibinfo {year} {2014}},\ \emph {\bibinfo
  {volume} {16}},\ \bibinfo {pages} {10345}}\BibitemShut {NoStop}%
\bibitem [{\citenamefont {Saltiel}\ \emph {et~al.}(1981)\citenamefont
  {Saltiel}, \citenamefont {March}, \citenamefont {Smothers}, \citenamefont
  {Stout},\ and\ \citenamefont {Charlton}}]{saltiel1981spin}%
  \BibitemOpen
  \bibfield  {author} {\bibinfo {author} {\bibfnamefont {J.}~\bibnamefont
  {Saltiel}}, \bibinfo {author} {\bibfnamefont {G.~R.}\ \bibnamefont {March}},
  \bibinfo {author} {\bibfnamefont {W.~K.}\ \bibnamefont {Smothers}}, \bibinfo
  {author} {\bibfnamefont {S.~A.}\ \bibnamefont {Stout}}, \ and\ \bibinfo
  {author} {\bibfnamefont {J.~L.}\ \bibnamefont {Charlton}},\ }\href@noop {}
  {\bibfield  {journal} {\bibinfo  {journal} {\emph {Journal of the American
  Chemical Society}}\ }\textbf {\bibinfo {year} {1981}},\ \emph {\bibinfo
  {volume} {103}},\ \bibinfo {pages} {7159}}\BibitemShut {NoStop}%
\bibitem [{\citenamefont {Tayebjee}\ \emph {et~al.}(2017)\citenamefont
  {Tayebjee}, \citenamefont {Sanders}, \citenamefont {Kumarasamy},
  \citenamefont {Campos}, \citenamefont {Sfeir},\ and\ \citenamefont
  {McCamey}}]{tayebjee2017quintet}%
  \BibitemOpen
  \bibfield  {author} {\bibinfo {author} {\bibfnamefont {M.~J.}\ \bibnamefont
  {Tayebjee}}, \bibinfo {author} {\bibfnamefont {S.~N.}\ \bibnamefont
  {Sanders}}, \bibinfo {author} {\bibfnamefont {E.}~\bibnamefont {Kumarasamy}},
  \bibinfo {author} {\bibfnamefont {L.~M.}\ \bibnamefont {Campos}}, \bibinfo
  {author} {\bibfnamefont {M.~Y.}\ \bibnamefont {Sfeir}}, \ and\ \bibinfo
  {author} {\bibfnamefont {D.~R.}\ \bibnamefont {McCamey}},\ }\href@noop {}
  {\bibfield  {journal} {\bibinfo  {journal} {\emph {Nature Physics}}\ }\textbf
  {\bibinfo {year} {2017}},\ \emph {\bibinfo {volume} {13}},\ \bibinfo {pages}
  {182}}\BibitemShut {NoStop}%
\bibitem [{\citenamefont {Calef}\ and\ \citenamefont
  {Deutch}(1983)}]{calef1983diffusion}%
  \BibitemOpen
  \bibfield  {author} {\bibinfo {author} {\bibfnamefont {D.~F.}\ \bibnamefont
  {Calef}}\ and\ \bibinfo {author} {\bibfnamefont {J.}~\bibnamefont {Deutch}},\
  }\href@noop {} {\bibfield  {journal} {\bibinfo  {journal} {\emph {Annual
  Review of Physical Chemistry}}\ }\textbf {\bibinfo {year} {1983}},\ \emph
  {\bibinfo {volume} {34}},\ \bibinfo {pages} {493}}\BibitemShut {NoStop}%
\bibitem [{\citenamefont {Aulin}\ \emph {et~al.}(2015)\citenamefont {Aulin},
  \citenamefont {van Sebille}, \citenamefont {Moes},\ and\ \citenamefont
  {Grozema}}]{aulin2015photochemical}%
  \BibitemOpen
  \bibfield  {author} {\bibinfo {author} {\bibfnamefont {Y.~V.}\ \bibnamefont
  {Aulin}}, \bibinfo {author} {\bibfnamefont {M.}~\bibnamefont {van Sebille}},
  \bibinfo {author} {\bibfnamefont {M.}~\bibnamefont {Moes}}, \ and\ \bibinfo
  {author} {\bibfnamefont {F.~C.}\ \bibnamefont {Grozema}},\ }\href@noop {}
  {\bibfield  {journal} {\bibinfo  {journal} {\emph {RSC Advances}}\ }\textbf
  {\bibinfo {year} {2015}},\ \emph {\bibinfo {volume} {5}},\ \bibinfo {pages}
  {107896}}\BibitemShut {NoStop}%
\bibitem [{\citenamefont {Th{\'e}venaz}\ \emph {et~al.}(2016)\citenamefont
  {Th{\'e}venaz}, \citenamefont {Monguzzi}, \citenamefont {Vanhecke},
  \citenamefont {Vadrucci}, \citenamefont {Meinardi}, \citenamefont {Simon},\
  and\ \citenamefont {Weder}}]{thevenaz2016thermoresponsive}%
  \BibitemOpen
  \bibfield  {author} {\bibinfo {author} {\bibfnamefont {D.~C.}\ \bibnamefont
  {Th{\'e}venaz}}, \bibinfo {author} {\bibfnamefont {A.}~\bibnamefont
  {Monguzzi}}, \bibinfo {author} {\bibfnamefont {D.}~\bibnamefont {Vanhecke}},
  \bibinfo {author} {\bibfnamefont {R.}~\bibnamefont {Vadrucci}}, \bibinfo
  {author} {\bibfnamefont {F.}~\bibnamefont {Meinardi}}, \bibinfo {author}
  {\bibfnamefont {Y.~C.}\ \bibnamefont {Simon}}, \ and\ \bibinfo {author}
  {\bibfnamefont {C.}~\bibnamefont {Weder}},\ }\href@noop {} {\bibfield
  {journal} {\bibinfo  {journal} {\emph {Materials Horizons}}\ }\textbf
  {\bibinfo {year} {2016}},\ \emph {\bibinfo {volume} {3}},\ \bibinfo {pages}
  {602}}\BibitemShut {NoStop}%
\bibitem [{\citenamefont {Schweitzer}\ and\ \citenamefont
  {Schmidt}(2003)}]{schweitzer2003physical}%
  \BibitemOpen
  \bibfield  {author} {\bibinfo {author} {\bibfnamefont {C.}~\bibnamefont
  {Schweitzer}}\ and\ \bibinfo {author} {\bibfnamefont {R.}~\bibnamefont
  {Schmidt}},\ }\href@noop {} {\bibfield  {journal} {\bibinfo  {journal} {\emph
  {Chemical Reviews}}\ }\textbf {\bibinfo {year} {2003}},\ \emph {\bibinfo
  {volume} {103}},\ \bibinfo {pages} {1685}}\BibitemShut {NoStop}%
\bibitem [{\citenamefont {Singh-Rachford}\ \emph {et~al.}(2009)\citenamefont
  {Singh-Rachford}, \citenamefont {Lott}, \citenamefont {Weder},\ and\
  \citenamefont {Castellano}}]{singh2009influence}%
  \BibitemOpen
  \bibfield  {author} {\bibinfo {author} {\bibfnamefont {T.~N.}\ \bibnamefont
  {Singh-Rachford}}, \bibinfo {author} {\bibfnamefont {J.}~\bibnamefont
  {Lott}}, \bibinfo {author} {\bibfnamefont {C.}~\bibnamefont {Weder}}, \ and\
  \bibinfo {author} {\bibfnamefont {F.~N.}\ \bibnamefont {Castellano}},\
  }\href@noop {} {\bibfield  {journal} {\bibinfo  {journal} {\emph {Journal of
  the American Chemical Society}}\ }\textbf {\bibinfo {year} {2009}},\ \emph
  {\bibinfo {volume} {131}},\ \bibinfo {pages} {12007}}\BibitemShut {NoStop}%
\bibitem [{\citenamefont {Askes}\ \emph {et~al.}(2017)\citenamefont {Askes},
  \citenamefont {Brodie}, \citenamefont {Bruylants},\ and\ \citenamefont
  {Bonnet}}]{askes2017temperature}%
  \BibitemOpen
  \bibfield  {author} {\bibinfo {author} {\bibfnamefont {S.~H.}\ \bibnamefont
  {Askes}}, \bibinfo {author} {\bibfnamefont {P.}~\bibnamefont {Brodie}},
  \bibinfo {author} {\bibfnamefont {G.}~\bibnamefont {Bruylants}}, \ and\
  \bibinfo {author} {\bibfnamefont {S.}~\bibnamefont {Bonnet}},\ }\href@noop {}
  {\bibfield  {journal} {\bibinfo  {journal} {\emph {The Journal of Physical
  Chemistry B}}\ }\textbf {\bibinfo {year} {2017}},\ \emph {\bibinfo {volume}
  {121}},\ \bibinfo {pages} {780}}\BibitemShut {NoStop}%
\bibitem [{\citenamefont {Massaro}\ \emph {et~al.}(2016)\citenamefont
  {Massaro}, \citenamefont {Hernando}, \citenamefont {Ruiz-Molina},
  \citenamefont {Roscini},\ and\ \citenamefont
  {Latterini}}]{massaro2016thermally}%
  \BibitemOpen
  \bibfield  {author} {\bibinfo {author} {\bibfnamefont {G.}~\bibnamefont
  {Massaro}}, \bibinfo {author} {\bibfnamefont {J.}~\bibnamefont {Hernando}},
  \bibinfo {author} {\bibfnamefont {D.}~\bibnamefont {Ruiz-Molina}}, \bibinfo
  {author} {\bibfnamefont {C.}~\bibnamefont {Roscini}}, \ and\ \bibinfo
  {author} {\bibfnamefont {L.}~\bibnamefont {Latterini}},\ }\href@noop {}
  {\bibfield  {journal} {\bibinfo  {journal} {\emph {Chemistry of Materials}}\
  }\textbf {\bibinfo {year} {2016}},\ \emph {\bibinfo {volume} {28}},\ \bibinfo
  {pages} {738}}\BibitemShut {NoStop}%
\bibitem [{\citenamefont {Xu}\ \emph {et~al.}(2018)\citenamefont {Xu},
  \citenamefont {Zou}, \citenamefont {Su}, \citenamefont {Yuan}, \citenamefont
  {Cao}, \citenamefont {Wang}, \citenamefont {Zhu}, \citenamefont {Feng},\ and\
  \citenamefont {Li}}]{xu2018ratiometric}%
  \BibitemOpen
  \bibfield  {author} {\bibinfo {author} {\bibfnamefont {M.}~\bibnamefont
  {Xu}}, \bibinfo {author} {\bibfnamefont {X.}~\bibnamefont {Zou}}, \bibinfo
  {author} {\bibfnamefont {Q.}~\bibnamefont {Su}}, \bibinfo {author}
  {\bibfnamefont {W.}~\bibnamefont {Yuan}}, \bibinfo {author} {\bibfnamefont
  {C.}~\bibnamefont {Cao}}, \bibinfo {author} {\bibfnamefont {Q.}~\bibnamefont
  {Wang}}, \bibinfo {author} {\bibfnamefont {X.}~\bibnamefont {Zhu}}, \bibinfo
  {author} {\bibfnamefont {W.}~\bibnamefont {Feng}}, \ and\ \bibinfo {author}
  {\bibfnamefont {F.}~\bibnamefont {Li}},\ }\href@noop {} {\bibfield  {journal}
  {\bibinfo  {journal} {\emph {Nature Communications}}\ }\textbf {\bibinfo
  {year} {2018}},\ \emph {\bibinfo {volume} {9}},\ \bibinfo {pages}
  {2698}}\BibitemShut {NoStop}%
\bibitem [{Note1()}]{Note1}%
  \BibitemOpen
  \bibinfo {note} {We omit the shape and temperature dependence of the LED
  spectrum throughout.}\BibitemShut {Stop}%
\bibitem [{\citenamefont {Xie}\ \emph {et~al.}(2011)\citenamefont {Xie},
  \citenamefont {Dai}, \citenamefont {Wang},\ and\ \citenamefont
  {Sumathy}}]{xie2011concentrated}%
  \BibitemOpen
  \bibfield  {author} {\bibinfo {author} {\bibfnamefont {W.}~\bibnamefont
  {Xie}}, \bibinfo {author} {\bibfnamefont {Y.}~\bibnamefont {Dai}}, \bibinfo
  {author} {\bibfnamefont {R.}~\bibnamefont {Wang}}, \ and\ \bibinfo {author}
  {\bibfnamefont {K.}~\bibnamefont {Sumathy}},\ }\href@noop {} {\bibfield
  {journal} {\bibinfo  {journal} {\emph {Renewable and Sustainable Energy
  Reviews}}\ }\textbf {\bibinfo {year} {2011}},\ \emph {\bibinfo {volume}
  {15}},\ \bibinfo {pages} {2588}}\BibitemShut {NoStop}%
\bibitem [{\citenamefont {Kalogirou}(2004)}]{kalogirou2004solar}%
  \BibitemOpen
  \bibfield  {author} {\bibinfo {author} {\bibfnamefont {S.~A.}\ \bibnamefont
  {Kalogirou}},\ }\href@noop {} {\bibfield  {journal} {\bibinfo  {journal}
  {\emph {Progress in Energy and Combustion Science}}\ }\textbf {\bibinfo
  {year} {2004}},\ \emph {\bibinfo {volume} {30}},\ \bibinfo {pages}
  {231}}\BibitemShut {NoStop}%
\bibitem [{\citenamefont {Badescu}(1995)}]{badescu1995radius}%
  \BibitemOpen
  \bibfield  {author} {\bibinfo {author} {\bibfnamefont {V.}~\bibnamefont
  {Badescu}},\ }\href@noop {} {\bibfield  {journal} {\bibinfo  {journal} {\emph
  {Acta Astronautica}}\ }\textbf {\bibinfo {year} {1995}},\ \emph {\bibinfo
  {volume} {36}},\ \bibinfo {pages} {135}}\BibitemShut {NoStop}%
\bibitem [{\citenamefont {Deng}\ \emph {et~al.}(2013)\citenamefont {Deng},
  \citenamefont {Sommer}, \citenamefont {Myahkostupov}, \citenamefont
  {Schanze},\ and\ \citenamefont {Castellano}}]{deng2013near}%
  \BibitemOpen
  \bibfield  {author} {\bibinfo {author} {\bibfnamefont {F.}~\bibnamefont
  {Deng}}, \bibinfo {author} {\bibfnamefont {J.~R.}\ \bibnamefont {Sommer}},
  \bibinfo {author} {\bibfnamefont {M.}~\bibnamefont {Myahkostupov}}, \bibinfo
  {author} {\bibfnamefont {K.~S.}\ \bibnamefont {Schanze}}, \ and\ \bibinfo
  {author} {\bibfnamefont {F.~N.}\ \bibnamefont {Castellano}},\ }\href@noop {}
  {\bibfield  {journal} {\bibinfo  {journal} {\emph {Chemical Communications}}\
  }\textbf {\bibinfo {year} {2013}},\ \emph {\bibinfo {volume} {49}},\ \bibinfo
  {pages} {7406}}\BibitemShut {NoStop}%
\bibitem [{\citenamefont {Zou}\ \emph {et~al.}(2012)\citenamefont {Zou},
  \citenamefont {Visser}, \citenamefont {Maduro}, \citenamefont
  {Pshenichnikov},\ and\ \citenamefont {Hummelen}}]{zou2012broadband}%
  \BibitemOpen
  \bibfield  {author} {\bibinfo {author} {\bibfnamefont {W.}~\bibnamefont
  {Zou}}, \bibinfo {author} {\bibfnamefont {C.}~\bibnamefont {Visser}},
  \bibinfo {author} {\bibfnamefont {J.~A.}\ \bibnamefont {Maduro}}, \bibinfo
  {author} {\bibfnamefont {M.~S.}\ \bibnamefont {Pshenichnikov}}, \ and\
  \bibinfo {author} {\bibfnamefont {J.~C.}\ \bibnamefont {Hummelen}},\
  }\href@noop {} {\bibfield  {journal} {\bibinfo  {journal} {\emph {Nature
  Photonics}}\ }\textbf {\bibinfo {year} {2012}},\ \emph {\bibinfo {volume}
  {6}},\ \bibinfo {pages} {560}}\BibitemShut {NoStop}%
\bibitem [{\citenamefont {Wu}\ \emph {et~al.}(2016)\citenamefont {Wu},
  \citenamefont {Congreve}, \citenamefont {Wilson}, \citenamefont {Jean},
  \citenamefont {Geva}, \citenamefont {Welborn}, \citenamefont {Van~Voorhis},
  \citenamefont {Bulovi{\'c}}, \citenamefont {Bawendi},\ and\ \citenamefont
  {Baldo}}]{wu2016solid}%
  \BibitemOpen
  \bibfield  {author} {\bibinfo {author} {\bibfnamefont {M.}~\bibnamefont
  {Wu}}, \bibinfo {author} {\bibfnamefont {D.~N.}\ \bibnamefont {Congreve}},
  \bibinfo {author} {\bibfnamefont {M.~W.}\ \bibnamefont {Wilson}}, \bibinfo
  {author} {\bibfnamefont {J.}~\bibnamefont {Jean}}, \bibinfo {author}
  {\bibfnamefont {N.}~\bibnamefont {Geva}}, \bibinfo {author} {\bibfnamefont
  {M.}~\bibnamefont {Welborn}}, \bibinfo {author} {\bibfnamefont
  {T.}~\bibnamefont {Van~Voorhis}}, \bibinfo {author} {\bibfnamefont
  {V.}~\bibnamefont {Bulovi{\'c}}}, \bibinfo {author} {\bibfnamefont {M.~G.}\
  \bibnamefont {Bawendi}}, \ and\ \bibinfo {author} {\bibfnamefont {M.~A.}\
  \bibnamefont {Baldo}},\ }\href@noop {} {\bibfield  {journal} {\bibinfo
  {journal} {\emph {Nature Photonics}}\ }\textbf {\bibinfo {year} {2016}},\
  \emph {\bibinfo {volume} {10}},\ \bibinfo {pages} {31}}\BibitemShut {NoStop}%
\bibitem [{\citenamefont {F\"uckel}\ \emph {et~al.}(2011)\citenamefont
  {F\"uckel}, \citenamefont {Roberts}, \citenamefont {Cheng}, \citenamefont
  {Clady}, \citenamefont {Piper}, \citenamefont {Ekins-Daukes}, \citenamefont
  {Crossley},\ and\ \citenamefont {Schmidt}}]{fuckel2011singlet}%
  \BibitemOpen
  \bibfield  {author} {\bibinfo {author} {\bibfnamefont {B.}~\bibnamefont
  {F\"uckel}}, \bibinfo {author} {\bibfnamefont {D.~A.}\ \bibnamefont
  {Roberts}}, \bibinfo {author} {\bibfnamefont {Y.~Y.}\ \bibnamefont {Cheng}},
  \bibinfo {author} {\bibfnamefont {R.~G.}\ \bibnamefont {Clady}}, \bibinfo
  {author} {\bibfnamefont {R.~B.}\ \bibnamefont {Piper}}, \bibinfo {author}
  {\bibfnamefont {N.}~\bibnamefont {Ekins-Daukes}}, \bibinfo {author}
  {\bibfnamefont {M.~J.}\ \bibnamefont {Crossley}}, \ and\ \bibinfo {author}
  {\bibfnamefont {T.~W.}\ \bibnamefont {Schmidt}},\ }\href@noop {} {\bibfield
  {journal} {\bibinfo  {journal} {\emph {The Journal of Physical Chemistry
  Letters}}\ }\textbf {\bibinfo {year} {2011}},\ \emph {\bibinfo {volume}
  {2}},\ \bibinfo {pages} {966}}\BibitemShut {NoStop}%
\bibitem [{\citenamefont {Yakutkin}\ \emph {et~al.}(2008)\citenamefont
  {Yakutkin}, \citenamefont {Aleshchenkov}, \citenamefont {Chernov},
  \citenamefont {Miteva}, \citenamefont {Nelles}, \citenamefont {Cheprakov},\
  and\ \citenamefont {Baluschev}}]{yakutkin2008towards}%
  \BibitemOpen
  \bibfield  {author} {\bibinfo {author} {\bibfnamefont {V.}~\bibnamefont
  {Yakutkin}}, \bibinfo {author} {\bibfnamefont {S.}~\bibnamefont
  {Aleshchenkov}}, \bibinfo {author} {\bibfnamefont {S.}~\bibnamefont
  {Chernov}}, \bibinfo {author} {\bibfnamefont {T.}~\bibnamefont {Miteva}},
  \bibinfo {author} {\bibfnamefont {G.}~\bibnamefont {Nelles}}, \bibinfo
  {author} {\bibfnamefont {A.}~\bibnamefont {Cheprakov}}, \ and\ \bibinfo
  {author} {\bibfnamefont {S.}~\bibnamefont {Baluschev}},\ }\href@noop {}
  {\bibfield  {journal} {\bibinfo  {journal} {\emph {Chemistry-A European
  Journal}}\ }\textbf {\bibinfo {year} {2008}},\ \emph {\bibinfo {volume}
  {14}},\ \bibinfo {pages} {9846}}\BibitemShut {NoStop}%
\bibitem [{\citenamefont {Singh-Rachford}\ \emph {et~al.}(2010)\citenamefont
  {Singh-Rachford}, \citenamefont {Nayak}, \citenamefont {Muro-Small},
  \citenamefont {Goeb}, \citenamefont {Therien},\ and\ \citenamefont
  {Castellano}}]{singh2010supermolecular}%
  \BibitemOpen
  \bibfield  {author} {\bibinfo {author} {\bibfnamefont {T.~N.}\ \bibnamefont
  {Singh-Rachford}}, \bibinfo {author} {\bibfnamefont {A.}~\bibnamefont
  {Nayak}}, \bibinfo {author} {\bibfnamefont {M.~L.}\ \bibnamefont
  {Muro-Small}}, \bibinfo {author} {\bibfnamefont {S.}~\bibnamefont {Goeb}},
  \bibinfo {author} {\bibfnamefont {M.~J.}\ \bibnamefont {Therien}}, \ and\
  \bibinfo {author} {\bibfnamefont {F.~N.}\ \bibnamefont {Castellano}},\
  }\href@noop {} {\bibfield  {journal} {\bibinfo  {journal} {\emph {Journal of
  the American Chemical Society}}\ }\textbf {\bibinfo {year} {2010}},\ \emph
  {\bibinfo {volume} {132}},\ \bibinfo {pages} {14203}}\BibitemShut {NoStop}%
\bibitem [{\citenamefont {Huang}\ \emph {et~al.}(2015)\citenamefont {Huang},
  \citenamefont {Li}, \citenamefont {Mahboub}, \citenamefont {Hanson},
  \citenamefont {Nichols}, \citenamefont {Le}, \citenamefont {Tang},\ and\
  \citenamefont {Bardeen}}]{huang2015hybrid}%
  \BibitemOpen
  \bibfield  {author} {\bibinfo {author} {\bibfnamefont {Z.}~\bibnamefont
  {Huang}}, \bibinfo {author} {\bibfnamefont {X.}~\bibnamefont {Li}}, \bibinfo
  {author} {\bibfnamefont {M.}~\bibnamefont {Mahboub}}, \bibinfo {author}
  {\bibfnamefont {K.~M.}\ \bibnamefont {Hanson}}, \bibinfo {author}
  {\bibfnamefont {V.~M.}\ \bibnamefont {Nichols}}, \bibinfo {author}
  {\bibfnamefont {H.}~\bibnamefont {Le}}, \bibinfo {author} {\bibfnamefont
  {M.~L.}\ \bibnamefont {Tang}}, \ and\ \bibinfo {author} {\bibfnamefont
  {C.~J.}\ \bibnamefont {Bardeen}},\ }\href@noop {} {\bibfield  {journal}
  {\bibinfo  {journal} {\emph {Nano Letters}}\ }\textbf {\bibinfo {year}
  {2015}},\ \emph {\bibinfo {volume} {15}},\ \bibinfo {pages}
  {5552}}\BibitemShut {NoStop}%
\bibitem [{\citenamefont {Iveland}\ \emph {et~al.}(2013)\citenamefont
  {Iveland}, \citenamefont {Martinelli}, \citenamefont {Peretti}, \citenamefont
  {Speck},\ and\ \citenamefont {Weisbuch}}]{iveland2013direct}%
  \BibitemOpen
  \bibfield  {author} {\bibinfo {author} {\bibfnamefont {J.}~\bibnamefont
  {Iveland}}, \bibinfo {author} {\bibfnamefont {L.}~\bibnamefont {Martinelli}},
  \bibinfo {author} {\bibfnamefont {J.}~\bibnamefont {Peretti}}, \bibinfo
  {author} {\bibfnamefont {J.~S.}\ \bibnamefont {Speck}}, \ and\ \bibinfo
  {author} {\bibfnamefont {C.}~\bibnamefont {Weisbuch}},\ }\href@noop {}
  {\bibfield  {journal} {\bibinfo  {journal} {\emph {Physical Review Letters}}\
  }\textbf {\bibinfo {year} {2013}},\ \emph {\bibinfo {volume} {110}},\
  \bibinfo {pages} {177406}}\BibitemShut {NoStop}%
\bibitem [{\citenamefont {Pozina}\ \emph {et~al.}(2015)\citenamefont {Pozina},
  \citenamefont {Ciechonski}, \citenamefont {Bi}, \citenamefont {Samuelson},\
  and\ \citenamefont {Monemar}}]{pozina2015dislocation}%
  \BibitemOpen
  \bibfield  {author} {\bibinfo {author} {\bibfnamefont {G.}~\bibnamefont
  {Pozina}}, \bibinfo {author} {\bibfnamefont {R.}~\bibnamefont {Ciechonski}},
  \bibinfo {author} {\bibfnamefont {Z.}~\bibnamefont {Bi}}, \bibinfo {author}
  {\bibfnamefont {L.}~\bibnamefont {Samuelson}}, \ and\ \bibinfo {author}
  {\bibfnamefont {B.}~\bibnamefont {Monemar}},\ }\href@noop {} {\bibfield
  {journal} {\bibinfo  {journal} {\emph {Applied Physics Letters}}\ }\textbf
  {\bibinfo {year} {2015}},\ \emph {\bibinfo {volume} {107}},\ \bibinfo {pages}
  {251106}}\BibitemShut {NoStop}%
\bibitem [{\citenamefont {Kioupakis}\ \emph {et~al.}(2011)\citenamefont
  {Kioupakis}, \citenamefont {Rinke}, \citenamefont {Delaney},\ and\
  \citenamefont {Van~de Walle}}]{kioupakis2011indirect}%
  \BibitemOpen
  \bibfield  {author} {\bibinfo {author} {\bibfnamefont {E.}~\bibnamefont
  {Kioupakis}}, \bibinfo {author} {\bibfnamefont {P.}~\bibnamefont {Rinke}},
  \bibinfo {author} {\bibfnamefont {K.~T.}\ \bibnamefont {Delaney}}, \ and\
  \bibinfo {author} {\bibfnamefont {C.~G.}\ \bibnamefont {Van~de Walle}},\
  }\href@noop {} {\bibfield  {journal} {\bibinfo  {journal} {\emph {Applied
  Physics Letters}}\ }\textbf {\bibinfo {year} {2011}},\ \emph {\bibinfo
  {volume} {98}},\ \bibinfo {pages} {161107}}\BibitemShut {NoStop}%
\bibitem [{\citenamefont {Binder}\ \emph {et~al.}(2013)\citenamefont {Binder},
  \citenamefont {Nirschl}, \citenamefont {Zeisel}, \citenamefont {Hager},
  \citenamefont {Lugauer}, \citenamefont {Sabathil}, \citenamefont {Bougeard},
  \citenamefont {Wagner},\ and\ \citenamefont
  {Galler}}]{binder2013identification}%
  \BibitemOpen
  \bibfield  {author} {\bibinfo {author} {\bibfnamefont {M.}~\bibnamefont
  {Binder}}, \bibinfo {author} {\bibfnamefont {A.}~\bibnamefont {Nirschl}},
  \bibinfo {author} {\bibfnamefont {R.}~\bibnamefont {Zeisel}}, \bibinfo
  {author} {\bibfnamefont {T.}~\bibnamefont {Hager}}, \bibinfo {author}
  {\bibfnamefont {H.-J.}\ \bibnamefont {Lugauer}}, \bibinfo {author}
  {\bibfnamefont {M.}~\bibnamefont {Sabathil}}, \bibinfo {author}
  {\bibfnamefont {D.}~\bibnamefont {Bougeard}}, \bibinfo {author}
  {\bibfnamefont {J.}~\bibnamefont {Wagner}}, \ and\ \bibinfo {author}
  {\bibfnamefont {B.}~\bibnamefont {Galler}},\ }\href@noop {} {\bibfield
  {journal} {\bibinfo  {journal} {\emph {Applied Physics Letters}}\ }\textbf
  {\bibinfo {year} {2013}},\ \emph {\bibinfo {volume} {103}},\ \bibinfo {pages}
  {071108}}\BibitemShut {NoStop}%
\bibitem [{\citenamefont {Verzellesi}\ \emph {et~al.}(2013)\citenamefont
  {Verzellesi}, \citenamefont {Saguatti}, \citenamefont {Meneghini},
  \citenamefont {Bertazzi}, \citenamefont {Goano}, \citenamefont {Meneghesso},\
  and\ \citenamefont {Zanoni}}]{verzellesi2013efficiency}%
  \BibitemOpen
  \bibfield  {author} {\bibinfo {author} {\bibfnamefont {G.}~\bibnamefont
  {Verzellesi}}, \bibinfo {author} {\bibfnamefont {D.}~\bibnamefont
  {Saguatti}}, \bibinfo {author} {\bibfnamefont {M.}~\bibnamefont {Meneghini}},
  \bibinfo {author} {\bibfnamefont {F.}~\bibnamefont {Bertazzi}}, \bibinfo
  {author} {\bibfnamefont {M.}~\bibnamefont {Goano}}, \bibinfo {author}
  {\bibfnamefont {G.}~\bibnamefont {Meneghesso}}, \ and\ \bibinfo {author}
  {\bibfnamefont {E.}~\bibnamefont {Zanoni}},\ }\href@noop {} {\bibfield
  {journal} {\bibinfo  {journal} {\emph {Journal of Applied Physics}}\ }\textbf
  {\bibinfo {year} {2013}},\ \emph {\bibinfo {volume} {114}},\ \bibinfo {pages}
  {071101}}\BibitemShut {NoStop}%
\bibitem [{\citenamefont {Cho}\ \emph {et~al.}(2013)\citenamefont {Cho},
  \citenamefont {Schubert},\ and\ \citenamefont {Kim}}]{cho2013efficiency}%
  \BibitemOpen
  \bibfield  {author} {\bibinfo {author} {\bibfnamefont {J.}~\bibnamefont
  {Cho}}, \bibinfo {author} {\bibfnamefont {E.~F.}\ \bibnamefont {Schubert}}, \
  and\ \bibinfo {author} {\bibfnamefont {J.~K.}\ \bibnamefont {Kim}},\
  }\href@noop {} {\bibfield  {journal} {\bibinfo  {journal} {\emph {Laser \&
  Photonics Reviews}}\ }\textbf {\bibinfo {year} {2013}},\ \emph {\bibinfo
  {volume} {7}},\ \bibinfo {pages} {408}}\BibitemShut {NoStop}%
\bibitem [{\citenamefont {Karpov}(2015)}]{karpov2015abc}%
  \BibitemOpen
  \bibfield  {author} {\bibinfo {author} {\bibfnamefont {S.}~\bibnamefont
  {Karpov}},\ }\href@noop {} {\bibfield  {journal} {\bibinfo  {journal} {\emph
  {Optical and Quantum Electronics}}\ }\textbf {\bibinfo {year} {2015}},\ \emph
  {\bibinfo {volume} {47}},\ \bibinfo {pages} {1293}}\BibitemShut {NoStop}%
\bibitem [{\citenamefont {Luo}\ \emph {et~al.}(2016)\citenamefont {Luo},
  \citenamefont {Hu}, \citenamefont {Liu},\ and\ \citenamefont
  {Wang}}]{luo2016heat}%
  \BibitemOpen
  \bibfield  {author} {\bibinfo {author} {\bibfnamefont {X.}~\bibnamefont
  {Luo}}, \bibinfo {author} {\bibfnamefont {R.}~\bibnamefont {Hu}}, \bibinfo
  {author} {\bibfnamefont {S.}~\bibnamefont {Liu}}, \ and\ \bibinfo {author}
  {\bibfnamefont {K.}~\bibnamefont {Wang}},\ }\href@noop {} {\bibfield
  {journal} {\bibinfo  {journal} {\emph {Progress in Energy and Combustion
  Science}}\ }\textbf {\bibinfo {year} {2016}},\ \emph {\bibinfo {volume}
  {56}},\ \bibinfo {pages} {1}}\BibitemShut {NoStop}%
\bibitem [{\citenamefont {Kuritzky}\ \emph {et~al.}(2017)\citenamefont
  {Kuritzky}, \citenamefont {Espenlaub}, \citenamefont {Yonkee}, \citenamefont
  {Pynn}, \citenamefont {DenBaars}, \citenamefont {Nakamura}, \citenamefont
  {Weisbuch},\ and\ \citenamefont {Speck}}]{kuritzky2017high}%
  \BibitemOpen
  \bibfield  {author} {\bibinfo {author} {\bibfnamefont {L.~Y.}\ \bibnamefont
  {Kuritzky}}, \bibinfo {author} {\bibfnamefont {A.~C.}\ \bibnamefont
  {Espenlaub}}, \bibinfo {author} {\bibfnamefont {B.~P.}\ \bibnamefont
  {Yonkee}}, \bibinfo {author} {\bibfnamefont {C.~D.}\ \bibnamefont {Pynn}},
  \bibinfo {author} {\bibfnamefont {S.~P.}\ \bibnamefont {DenBaars}}, \bibinfo
  {author} {\bibfnamefont {S.}~\bibnamefont {Nakamura}}, \bibinfo {author}
  {\bibfnamefont {C.}~\bibnamefont {Weisbuch}}, \ and\ \bibinfo {author}
  {\bibfnamefont {J.~S.}\ \bibnamefont {Speck}},\ }\href@noop {} {\bibfield
  {journal} {\bibinfo  {journal} {\emph {Optics Express}}\ }\textbf {\bibinfo
  {year} {2017}},\ \emph {\bibinfo {volume} {25}},\ \bibinfo {pages}
  {30696}}\BibitemShut {NoStop}%
\bibitem [{\citenamefont {Stanley}\ \emph {et~al.}(1994)\citenamefont
  {Stanley}, \citenamefont {Houdre}, \citenamefont {Oesterle}, \citenamefont
  {Gailhanou},\ and\ \citenamefont {Ilegems}}]{stanley1994ultrahigh}%
  \BibitemOpen
  \bibfield  {author} {\bibinfo {author} {\bibfnamefont {R.}~\bibnamefont
  {Stanley}}, \bibinfo {author} {\bibfnamefont {R.}~\bibnamefont {Houdre}},
  \bibinfo {author} {\bibfnamefont {U.}~\bibnamefont {Oesterle}}, \bibinfo
  {author} {\bibfnamefont {M.}~\bibnamefont {Gailhanou}}, \ and\ \bibinfo
  {author} {\bibfnamefont {M.}~\bibnamefont {Ilegems}},\ }\href@noop {}
  {\bibfield  {journal} {\bibinfo  {journal} {\emph {Applied Physics Letters}}\
  }\textbf {\bibinfo {year} {1994}},\ \emph {\bibinfo {volume} {65}},\ \bibinfo
  {pages} {1883}}\BibitemShut {NoStop}%
\bibitem [{\citenamefont {Schulz}(1989)}]{schulz1989wavelength}%
  \BibitemOpen
  \bibfield  {author} {\bibinfo {author} {\bibfnamefont {P.}~\bibnamefont
  {Schulz}},\ }\href@noop {} {\bibfield  {journal} {\bibinfo  {journal} {\emph
  {Applied Optics}}\ }\textbf {\bibinfo {year} {1989}},\ \emph {\bibinfo
  {volume} {28}},\ \bibinfo {pages} {4458}}\BibitemShut {NoStop}%
\bibitem [{\citenamefont {Gauthier}\ \emph {et~al.}(1986)\citenamefont
  {Gauthier}, \citenamefont {Narum},\ and\ \citenamefont
  {Boyd}}]{gauthier1986simple}%
  \BibitemOpen
  \bibfield  {author} {\bibinfo {author} {\bibfnamefont {D.~J.}\ \bibnamefont
  {Gauthier}}, \bibinfo {author} {\bibfnamefont {P.}~\bibnamefont {Narum}}, \
  and\ \bibinfo {author} {\bibfnamefont {R.~W.}\ \bibnamefont {Boyd}},\
  }\href@noop {} {\bibfield  {journal} {\bibinfo  {journal} {\emph {Optics
  Letters}}\ }\textbf {\bibinfo {year} {1986}},\ \emph {\bibinfo {volume}
  {11}},\ \bibinfo {pages} {623}}\BibitemShut {NoStop}%
\bibitem [{\citenamefont {Cheng}\ \emph {et~al.}(2012)\citenamefont {Cheng},
  \citenamefont {Huang},\ and\ \citenamefont {Lin}}]{cheng2012heat}%
  \BibitemOpen
  \bibfield  {author} {\bibinfo {author} {\bibfnamefont {H.~H.}\ \bibnamefont
  {Cheng}}, \bibinfo {author} {\bibfnamefont {D.-S.}\ \bibnamefont {Huang}}, \
  and\ \bibinfo {author} {\bibfnamefont {M.-T.}\ \bibnamefont {Lin}},\
  }\href@noop {} {\bibfield  {journal} {\bibinfo  {journal} {\emph
  {Microelectronics Reliability}}\ }\textbf {\bibinfo {year} {2012}},\ \emph
  {\bibinfo {volume} {52}},\ \bibinfo {pages} {905}}\BibitemShut {NoStop}%
\bibitem [{\citenamefont {Lin}\ \emph {et~al.}(2011)\citenamefont {Lin},
  \citenamefont {Wang}, \citenamefont {Huo}, \citenamefont {Hu}, \citenamefont
  {Chen}, \citenamefont {Zhang},\ and\ \citenamefont {Lee}}]{lin2011heat}%
  \BibitemOpen
  \bibfield  {author} {\bibinfo {author} {\bibfnamefont {Z.}~\bibnamefont
  {Lin}}, \bibinfo {author} {\bibfnamefont {S.}~\bibnamefont {Wang}}, \bibinfo
  {author} {\bibfnamefont {J.}~\bibnamefont {Huo}}, \bibinfo {author}
  {\bibfnamefont {Y.}~\bibnamefont {Hu}}, \bibinfo {author} {\bibfnamefont
  {J.}~\bibnamefont {Chen}}, \bibinfo {author} {\bibfnamefont {W.}~\bibnamefont
  {Zhang}}, \ and\ \bibinfo {author} {\bibfnamefont {E.}~\bibnamefont {Lee}},\
  }\href@noop {} {\bibfield  {journal} {\bibinfo  {journal} {\emph {Applied
  Thermal Engineering}}\ }\textbf {\bibinfo {year} {2011}},\ \emph {\bibinfo
  {volume} {31}},\ \bibinfo {pages} {2221}}\BibitemShut {NoStop}%
\bibitem [{\citenamefont {Weng}(2009)}]{weng2009advanced}%
  \BibitemOpen
  \bibfield  {author} {\bibinfo {author} {\bibfnamefont {C.-J.}\ \bibnamefont
  {Weng}},\ }\href@noop {} {\bibfield  {journal} {\bibinfo  {journal} {\emph
  {International Communications in Heat and Mass Transfer}}\ }\textbf {\bibinfo
  {year} {2009}},\ \emph {\bibinfo {volume} {36}},\ \bibinfo {pages}
  {245}}\BibitemShut {NoStop}%
\bibitem [{\citenamefont {Yung}\ \emph
  {et~al.}(2014{\natexlab{a}})\citenamefont {Yung}, \citenamefont {Liem},
  \citenamefont {Choy},\ and\ \citenamefont {Cai}}]{yung2014thermal}%
  \BibitemOpen
  \bibfield  {author} {\bibinfo {author} {\bibfnamefont {K.}~\bibnamefont
  {Yung}}, \bibinfo {author} {\bibfnamefont {H.}~\bibnamefont {Liem}}, \bibinfo
  {author} {\bibfnamefont {H.}~\bibnamefont {Choy}}, \ and\ \bibinfo {author}
  {\bibfnamefont {Z.}~\bibnamefont {Cai}},\ }\href@noop {} {\bibfield
  {journal} {\bibinfo  {journal} {\emph {Applied Thermal Engineering}}\
  }\textbf {\bibinfo {year} {2014}{\natexlab{a}}},\ \emph {\bibinfo {volume}
  {63}},\ \bibinfo {pages} {105}}\BibitemShut {NoStop}%
\bibitem [{\citenamefont {Yung}\ \emph
  {et~al.}(2014{\natexlab{b}})\citenamefont {Yung}, \citenamefont {Liem},\ and\
  \citenamefont {Choy}}]{yung2014heat}%
  \BibitemOpen
  \bibfield  {author} {\bibinfo {author} {\bibfnamefont {K.}~\bibnamefont
  {Yung}}, \bibinfo {author} {\bibfnamefont {H.}~\bibnamefont {Liem}}, \ and\
  \bibinfo {author} {\bibfnamefont {H.}~\bibnamefont {Choy}},\ }\href@noop {}
  {\bibfield  {journal} {\bibinfo  {journal} {\emph {International
  Communications in Heat and Mass Transfer}}\ }\textbf {\bibinfo {year}
  {2014}{\natexlab{b}}},\ \emph {\bibinfo {volume} {53}},\ \bibinfo {pages}
  {79}}\BibitemShut {NoStop}%
\bibitem [{\citenamefont {Suzuki}\ and\ \citenamefont
  {Matsumoto}(2008)}]{suzuki2008lithography}%
  \BibitemOpen
  \bibfield  {author} {\bibinfo {author} {\bibfnamefont {S.}~\bibnamefont
  {Suzuki}}\ and\ \bibinfo {author} {\bibfnamefont {Y.}~\bibnamefont
  {Matsumoto}},\ }\href@noop {} {\bibfield  {journal} {\bibinfo  {journal}
  {\emph {Microsystem Technologies}}\ }\textbf {\bibinfo {year} {2008}},\ \emph
  {\bibinfo {volume} {14}},\ \bibinfo {pages} {1291}}\BibitemShut {NoStop}%
\bibitem [{\citenamefont {Harburg}\ \emph {et~al.}(2013)\citenamefont
  {Harburg}, \citenamefont {Khan}, \citenamefont {Herrault}, \citenamefont
  {Kim}, \citenamefont {Levey},\ and\ \citenamefont
  {Sullivan}}]{harburg2013chip}%
  \BibitemOpen
  \bibfield  {author} {\bibinfo {author} {\bibfnamefont {D.~V.}\ \bibnamefont
  {Harburg}}, \bibinfo {author} {\bibfnamefont {G.~R.}\ \bibnamefont {Khan}},
  \bibinfo {author} {\bibfnamefont {F.}~\bibnamefont {Herrault}}, \bibinfo
  {author} {\bibfnamefont {J.}~\bibnamefont {Kim}}, \bibinfo {author}
  {\bibfnamefont {C.~G.}\ \bibnamefont {Levey}}, \ and\ \bibinfo {author}
  {\bibfnamefont {C.~R.}\ \bibnamefont {Sullivan}},\ }in\ \href@noop {} {\emph
  {\bibinfo {booktitle} {Solid-State Sensors, Actuators and Microsystems, 2013
  Transducers \& Eurosensors XXVII: The 17th International Conference on}}}\
  (\bibinfo {organization} {IEEE},\ \bibinfo {year} {2013})\ pp.\ \bibinfo
  {pages} {701--704}\BibitemShut {NoStop}%
\bibitem [{\citenamefont {Guijt}\ and\ \citenamefont
  {Breadmore}(2008)}]{guijt2008maskless}%
  \BibitemOpen
  \bibfield  {author} {\bibinfo {author} {\bibfnamefont {R.~M.}\ \bibnamefont
  {Guijt}}\ and\ \bibinfo {author} {\bibfnamefont {M.~C.}\ \bibnamefont
  {Breadmore}},\ }\href@noop {} {\bibfield  {journal} {\bibinfo  {journal}
  {\emph {Lab on a Chip}}\ }\textbf {\bibinfo {year} {2008}},\ \emph {\bibinfo
  {volume} {8}},\ \bibinfo {pages} {1402}}\BibitemShut {NoStop}%
\bibitem [{\citenamefont {Kim}\ and\ \citenamefont
  {Kim}(2012)}]{kim2012encapsulated}%
  \BibitemOpen
  \bibfield  {author} {\bibinfo {author} {\bibfnamefont {J.-H.}\ \bibnamefont
  {Kim}}\ and\ \bibinfo {author} {\bibfnamefont {J.-H.}\ \bibnamefont {Kim}},\
  }\href@noop {} {\bibfield  {journal} {\bibinfo  {journal} {\emph {Journal of
  the American Chemical Society}}\ }\textbf {\bibinfo {year} {2012}},\ \emph
  {\bibinfo {volume} {134}},\ \bibinfo {pages} {17478}}\BibitemShut {NoStop}%
\bibitem [{\citenamefont {Wu}\ \emph {et~al.}(2013)\citenamefont {Wu},
  \citenamefont {Sun}, \citenamefont {Cui},\ and\ \citenamefont
  {Zhao}}]{wu2013observation}%
  \BibitemOpen
  \bibfield  {author} {\bibinfo {author} {\bibfnamefont {W.}~\bibnamefont
  {Wu}}, \bibinfo {author} {\bibfnamefont {J.}~\bibnamefont {Sun}}, \bibinfo
  {author} {\bibfnamefont {X.}~\bibnamefont {Cui}}, \ and\ \bibinfo {author}
  {\bibfnamefont {J.}~\bibnamefont {Zhao}},\ }\href@noop {} {\bibfield
  {journal} {\bibinfo  {journal} {\emph {Journal of Materials Chemistry C}}\
  }\textbf {\bibinfo {year} {2013}},\ \emph {\bibinfo {volume} {1}},\ \bibinfo
  {pages} {4577}}\BibitemShut {NoStop}%
\end{thebibliography}%

\end{document}